\documentclass[showpacs,preprintnumbers,amsmath,amssymb,floatfix,
nofootinbib,aps]{revtex4}

\usepackage[dvips]{graphicx,graphics}
\usepackage{dcolumn}
\usepackage{bm}
\usepackage{epsfig}

\newcommand{\dirac}{\partial\llap{$\diagup$\kern-2pt}}

\begin{document}

\title{Vector interaction, charge neutrality and multiple chiral critical point structures}
\author{Zhao Zhang}\email{zhaozhang@pku.org.cn}
\author{Teiji Kunihiro}\email{kunihiro@ruby.scphys.kyoto-u.ac.jp}
\affiliation{Department of Physics, Kyoto University, Kyoto
606-8502, Japan}


\begin{abstract}

We investigate the combined effect of the repulsive vector
interaction and the positive electric chemical potential on the
chiral phase transition by considering neutral color
superconductivity. The chiral condensate, diquark condensate and
quark number densities are solved in both two-flavor and
two-plus-one-flavor Nambu-Jona-Lasinio models with the so called
Kobayashi-Maskawa-'t Hooft term under the charge-neutrality
constraint. We demonstrate that multiple chiral critical-point
structures always exist in the Nambu-Jona-Lasinio model within the
self-consistent mean-field approximation, and that the number of
chiral critical points can vary from zero to four, which is
dependent on the magnitudes of vector interaction and the diquark
coupling. The difference between the dynamical chemical potentials
induced by vector interaction for u and d quarks can effectively
reduce the Fermi sphere disparity between the two flavors of
diquark  pairing. Thus the vector interaction works to
significantly suppress the unstable region associated with
chromomagnetic instability in the phase of neutral asymmetric
homogeneous color superconductivity.

\end{abstract}
\preprint{KUNS-2201} \maketitle


\section{INTRODUCTION}
  It is generally believed that QCD
exhibits a rich phase structure in an extreme environment such as
at high temperature and high baryon chemical potential. For the
chiral phase transition, it is a widely accepted view that the
critical point(s) (i.e.\ the end point of the first-order phase
boundary) should exist at finite temperature and density.
Recently, a schematic $T$-$\mu$ phase diagram with one critical
point was extensively adopted in the
literature\cite{Asakawa:1989bq,Barducci:1989,Stephanov:2007fk}.

  In the last decade the color-superconducting (CSC) phase has
attracted a lot of theoretical interest and prompted extensive
studies of dense and cold quark
matter~\cite{WilczekReview,RischkeReview,BuballaReview,AlfordReview}.
At asymptotically high density that justifies the perturbative QCD
calculations, the color-flavor locked (CFL) phase~\cite{CFL} has
been established as the ground state of quark matter. For the
intermediate density region which may exist in the core of a
compact star, the nonperturbative features of QCD play a more
important role in the phase structure of QCD, and both CFL and
non-CFL CSC phases may appear in this region.

  The possible emergence of a CFL or a non-CFL CSC phase around the chiral
transition boundary at low temperatures should
affect the chiral phase transition,  and hence the interplay
between the chiral condensate and the diquark condensate may
result in an unexpected
phase structure of QCD. Such an example was first presented
in~\cite{KitazawaVector} in a two-flavor Nambu-Jona-Lasinio (NJL)
model with the vector interaction included: It was found that the
repulsive vector interaction can lead to a two-critical-point
structure in the $T$-$\mu$ phase diagram of QCD, which is due to
the fact that the density-density correlation induced by the
vector interaction effectively enhances the competition between
the chiral and diquark correlations while weakening the
first-order chiral
restoration\cite{Asakawa:1989bq,Klimt,Buballa1996}.
We note that the renormalization-group analysis
~\cite{ref:EHS,ref:SW-reno}, the chiral
instanton$\_$anti-instanton molecule model ~\cite{ref:SS}, and the
truncated Dyson-Schwinger model of QCD ~\cite{ref:RWP} all support
the existence of the vector-vector four-quark interaction. The
vector-vector interaction
 may be responsible for the vacuum properties of vector mesons
in low-energy effective theories of
QCD\cite{Ebert:1983,3-NJL1,Klimt:1989pm}.

In a quite different context, a realization of a similar
two-critical-point structure of the QCD phase diagram
 has been recently conjectured
~\cite{Hatsuda:2006ps}:  That is,
 the $\mathrm{U_A}(1)$-breaking
vertex may induce a new critical point.
This conjecture is based on a general Ginzburg-Landau
theory constrained by QCD symmetries. There, the three-flavor
anomaly term generates the cubic coupling between the chiral
 and diquark condensates. It has been argued that
the resultant crossover of chiral restoration at small
temperatures embodies the hadron-quark continuity
hypothesis~\cite{Continuity}. Further investigation is necessary
to clarify whether the new critical point really  exists or not
with reasonable parameter.

Beside the above two cases, another mechanism  which can induce
multiple critical-point structure was demonstrated in
~\cite{Zhang:2008wx}, where it was first disclosed that the
positive electric chemical potential $\mu_e$ required by the
charge-neutrality constraint plays a similar role as the repulsive
vector interaction on the chiral phase transition in a four-quark
interaction model: In a simple two-flavor NJL model, the same
two-critical-point structure found in ~\cite{KitazawaVector} was
obtained. In addition, positive $\mu_e$ also represents  the
magnitude of disparity between Fermi spheres of u and d quarks
when taking into account the local charge-neutrality constraint.
For an asymmetric homogeneous system including two different
flavor quarks with the mismatched Fermi spheres,
 the energy gap of the Cooper
pairing between these two-flavor quarks can increase with
temperature. This unusual behavior of the energy gap
 is due to the smearing of the Fermi surface by
temperature. For the two-flavor neutral CSC phase, this
unconventional thermal behavior of the diquark condensate can lead
to a special competition between the chiral condensate and the
diquark condensate, which may be enhanced with increasing
temperature. For some model parameter regions, this abnormal
competition induced by $\mu_e$ can result in a nontrivial
three-critical-point phase structure~\cite{Zhang:2008wx}.
The competition of two order parameters with the external
constraint(s) should also suggest a general mechanism for a
realization of a multiple critical-point structure, which may have
some implications to study the phase transitions in condensed
matter physics.

   Note, however, that  a very simple two-flavor
NJL model with only scalar quark-antiquark and diquark interaction
 was used in Ref.~\cite{Zhang:2008wx}:
 If the model produces a relatively large vacuum
quark mass, no multiple critical points appear in such formalism
even though the charge-neutral constraint is taken into account
for the CSC phase. For a more realistic situation, both the vector
interaction and the strange quark degree of freedom should  be
taken into account.

  Because in the chiral phase transition similar roles are played by
the repulsive vector interaction and the electric chemical
potential under the neutrality constraint, one can expect that the
chiral restoration will be weakened more significantly when
simultaneously taking into account both the ingredients. In this
case, it is possible that the multiple critical point structures
shown in \cite{Zhang:2008wx} will become more robust, which we
will  show in the present work. In addition, for the neutral quark
matter system, because of the quark density discrepancy, the
vector interaction should also give different contributions to the
dynamical chemical potentials for u and d quarks. This may have
important influence on the property of the asymmetric homogeneous
CSC phase which suffers from the so called chromomagnetic
instability characterized by the negative Meissner mass squared
\cite{Huang:2004bg}. This is a significant observation and will be
shown to be the case in the following sections.

 There have been extensive studies within the NJL model on the phase diagram of dense and locally neutral three-flavor
quark matter with strange quarks explicitly included
\cite{Ruester:2005jc,Abuki:2005ms}: In the strong coupling case,
the chiral breaking phase at low temperature is bordered by the
2CSC phase,  while for the weak coupling case it is surrounded by
both normal quark matter and a gapless phase called g2CSC.
However, no new critical point was found in
~\cite{Ruester:2005jc,Abuki:2005ms}, in contrast to the result
reported in ~\cite{Zhang:2008wx}. Notice, however,
 that the vector interaction  was not included in
~\cite{Ruester:2005jc,Abuki:2005ms}, and hence only a strong
first-order chiral restoration happened in the low temperature
region.
We can expect that the qualitative feature of  the phase diagram,
even for the three-flavor quark matter reported in
~\cite{Ruester:2005jc,Abuki:2005ms}, may change when the repulsive
vector interaction is taken into account together with the
charge-neutrality constraint. Again, we shall see that this is
actually the case,  and there appear multiple critical points in
the phase diagram of three-flavor quark matter.

 The paper is organized as follows. In the next section, a nonlocal two-flavor
NJL model is used to investigate the $(T,\mu)$ phase structure of
QCD, with both vector interaction and charge-neutrality being
taken into account. The extension of the work to the
two-plus-one-flavor case with the so-called Kobayashi-Maskawa-'t
Hooft(KMT) term\cite{KMT}is presented in Sec.III. The final
section is devoted to a summary and concluding remarks.


\section{Two-flavor case}
\label{sec: Two flavor }

We will demonstrate the joint effect of the vector interaction and
neutral electric-charge constraint on the QCD phase diagram within
a nonlocal two-flavor NJL model.


\subsection{Model}

NJL-type models have been extensively used to investigate the CSC
phase transition at moderate and large
densities~\cite{BuballaReview}: See the previous study without the
CSC~\cite{Hatsuda:1994,Asakawa:1989bq,Klimt,Klevansky:1992}. The
advantage of a NJL model is that it can investigate the interplay
between the chiral condensate, the diquark condensate and the
quark number density on the same footing. For the two-flavor case,
a nonlocal NJL model
\cite{Alford:1997zt,Schmidt:1994di,Bowler:1994ir,Blaschke:2000gd,GomezDumm:2005hy,Aguilera:2006cj,Grigorian:2006qe}
is adopted here, which has the Lagrangian density,
\begin{eqnarray}
\cal{L} &=& \bar\psi \, ( i \dirac - \hat{m} \, ) \psi +G_S \left[
\left( \bar{q}(x) q(x) \right)^2 + \left( \bar {q}(x) i \gamma_5
\vec{\tau} q(x) \right)^2 \right]-G_V \sum_{i=0}^3\left[ \left(
\bar{q}(x) \gamma^\mu \tau_i q(x) \right)^2 + \left( \bar q(x) i
\gamma^\mu \gamma_5 \tau_i q(x) \right)^2 \right]
\nonumber \\
&+& G_D \sum_{A} \left[\bar{q}(x) \gamma_5 \tau_2\lambda_A q_C(x)
\right] \left[ \bar{q}_C(x)\gamma_5\tau_2\lambda_A q(x) \right],
 \,
\label{Lagrangian2f}
\end{eqnarray}
where
\begin{eqnarray}
q(x)=\int{dy^4\tilde{f}(x-y)\psi(y)}\, ,
 \,\,\,q_C(x)=\int{dy^4\tilde{f}(x-y)\psi_C(y)}\,,\,\,
 \psi_C=C\bar{\psi}^T
\end{eqnarray}
and $C=i\gamma_0\gamma_2$ stands for the Dirac charge conjugation
matrix. The three coupling constants, namely $G_S$, $G_V$, and
$G_D$,  belong to the scalar mesonic channel,  the vector mesonic
channel, and the diquark channel, respectively. The current quark
mass matrix is given by $\widehat{m}=\text{diag}(m_u,m_d)$ in two
flavors,  and we shall work in the isospin symmetric limit with
$m_u=m_d=m$. We note that $\lambda_A$'s are the antisymmetric
Gell-Mann matrices (i.e.\ $A$ runs over $ 2, 5, 7$ only) for the
color SU(3) group,  while $\tau_0$  and the $\vec{\tau}$'s are the
unit matrix and Pauli matrices in flavor space, respectively. In
contrast to the scalar interaction, the vector part has
$U(2){\times}U(2)$ flavor symmetry,   and hence the vector terms
can be decomposed into
$\sim(\bar{u}\gamma^{\mu}u)^2\,+\,(\bar{d}\gamma^{\mu}d)^2$
without the flavor mixing term like
$\bar{u}\gamma^{\mu}u\bar{d}\gamma_{\mu}d$, although there are
terms like $\bar{u}\gamma^{\mu}d\bar{d}\gamma_{\mu}u$.

In contrast to the local NJL model,
Lagrangian~(\ref{Lagrangian2f}) is formulated with a nonlocal
interaction, which is controlled by a form factor $\tilde{f}(x)$.
This type of model can be considered as a special case of the
truncated Dyson-Schwinger equation  with a separable effective
gluon propagator or an instantaneous nonlocal chiral quark model.
The main purpose of adopting such a nonlocal interaction model in
our study is to deal with the one-loop ultraviolet divergence,
especially for the calculation of the Meissner mass squared at
finite temperature \cite{Kiriyama:2006jp}. For convenience, we
follow the model parametrization in \cite{Grigorian:2006qe} and
use the so-called Lorentzian form factor with the form
\begin{equation}
f^2(p=|{\mathbf{p}}|)=g(p)=\frac{1}{1 + (\frac{p}{\Lambda})^{2a}},
\label{formfactor}
\end{equation}
where $f(p)$ is the Fourier transformation of the form factor
$\tilde{f}(x)$. In Eq.(\ref{formfactor}), $a$ is a dimensionless
parameter and $\Lambda$ stands for the scale parameter of the
model. To check the sensitivity of the main results on the model
parameter choice, three sets of model parameters are adopted in
this paper, which are listed in Tabel.(\ref{tab1})\footnote{ We
mention that our results obtained in the Lorentzian nonlocal
cutoff scheme are equally obtained even with the traditional sharp
cutoff scheme. In this sense, our results, especially the
realization of the multiple critical-point structures of the phase
diagram are robust and not artificial.} By choosing $a=10$
\footnote{We have checked the main conclusion of this paper is
insensitive to a or the form of $g(p)$ if it can give reasonable
vacuum properties of QCD.}, the other three model parameters,
namely $\Lambda$, $G_S$ and $m$, are determined by the vacuum
physical quantities of the pion mass $M_\pi=135\text{MeV}$, the
pion decay constant $f_\pi=92.4 \text{MeV}$,  and the quark
condensate $-\langle \bar{u}u\rangle ^{1/3}\approx {250}
\text{MeV}$. Note that the value of the current quark mass is
around ${5}\text{MeV}$ and the Gell-Mann-Oakes-Renner relation
 holds well for these parametrizations of the model
parameters~\cite{Grigorian:2006qe}.

\begin{table}{
\begin{tabular}{c|c|c|c|c|c}
\hline {Parameter set  }&  \quad {\quad m(MeV) \quad}&
  \quad {\quad $G_S\Lambda^2$ \quad}&
  \quad {\quad $\Lambda$ (MeV)} \quad&
{$- \langle \bar{u}u\rangle ^{1/3}$ (MeV)}&
{$M(p=0)$ (MeV)}\\
\hline\hline Set 1 & 5.01137 & 2.64310  & 600.271  & 248.440 & 400
\\
\hline Set 2 & 4.92671 & 2.51088  & 617.968  & 249.906 & 367.5
\\
\hline Set 3 & 4.70805 & 2.35908  & 649.168  & 253.699 & 330
\\
\hline\hline
\end{tabular}
\caption{Model parameter sets for two-flavor nonlocal NJL.}
\label{tab1}}
\end{table}

In the nonlocal NJL model, the produced constituent quark mass is
momentum dependent. Table (\ref{tab1}) shows that the vacuum
constituent quark mass $M(p=0)$ ranges from $330\, \text{MeV}$ to
$400\, \text{MeV}$ for the listed three sets of model parameters
which almost reproduce the same vacuum physical quantities.
Although the vacuum constituent quark mass is not an observable,
all the values of $M(p=0)$ in Table (\ref{tab1}) are
phenomenologically acceptable because  the mass $3M(p=0)$ is
larger than the nucleon mass in the vacuum, which is a bound state
of three quarks.

Because there are  no reliable constraints on $G_V$ and $G_D$
within the two-flavor NJL model, these two model parameters are
taken as free parameters in our treatment. The standard ratios of
$G_V/G_S$ and $G_D/G_S$ from Fierz transformation based on the
local color current-current interaction are 0.5 and 0.75,
respectively.
 In addition, in the molecular instanton liquid model, the ratio $G_V/G_S$ is about 0.25.
  Combining this point with the Fierz transformation, the reasonable value of  $G_V/G_S$ may
be located in the range from 0 to 0.5.
 In the
following numerical calculations, we will focus on the influence
of the vector interaction on the phase diagram by varying
$G_V/G_S$ while fixing $G_D/G_S$ as its standard value. In the
literature, such a choice of diquark interaction is usually called
the intermediate coupling.


\subsection{Thermodynamic Potential with Neutrality Condition}

The grand partition function is given by
\begin{equation}
Z\equiv{e^{-{\Omega}V/T}}=\int{D\bar{\psi}D\psi}
e^{i\int{dx^4}(\cal{L}+{\psi^\dag}\hat{\mu}\psi)},
\end{equation}
where $\Omega$ is the thermodynamic potential density and
$\hat{\mu}$ is the quark chemical-potential matrix. In general,
the quark chemical-potential matrix $\hat{\mu}$ takes the
form~\cite{Alford:2002kj}
\begin{equation}
 \hat{\mu} = \mu - \mu_e Q + \mu_3T_3 + \mu_8T_8,
\end{equation}
where $\mu$ is the quark chemical potential (i.e.\ one third of
the baryon chemical potential), $\mu_e$ is the chemical potential
associated with the (negative) electric-charge, and $\mu_3$ and
$\mu_8$ represent the color chemical potentials corresponding to
the Cartan subalgebra in color SU(3) space.  The explicit form of
the electric charge matrix is
$Q=\text{diag}(\frac{2}{3},-\frac{1}{3})$ in flavor space, and the
color charge matrices are
$T_3=\text{diag}(\frac{1}{2},-\frac{1}{2},0)$ and
$T_8=\text{diag}(\frac{1}{3},\frac{1}{3}, -\frac{2}{3})$ in color
space.  The chemical potentials for different quarks are listed
below:
\begin{equation}
 \begin{split}
&\mu_{ru} = \mu-\tfrac{2}{3}\mu_e+\tfrac{1}{2}\mu_3+\tfrac{1}{3}\mu_8 \,,
 \qquad
 \mu_{gu} = \mu-\tfrac{2}{3}\mu_e-\tfrac{1}{2}\mu_3+\tfrac{1}{3}\mu_8 \,,\\
&\mu_{rd} = \mu+\tfrac{1}{3}\mu_e+\tfrac{1}{2}\mu_3+\tfrac{1}{3}\mu_8 \,,
 \qquad
 \mu_{gd} = \mu+\tfrac{1}{3}\mu_e-\tfrac{1}{2}\mu_3+\tfrac{1}{3}\mu_8 \,,\\
&\mu_{bu} = \mu-\tfrac{2}{3}\mu_e-\tfrac{2}{3}\mu_8 \,, \qquad\qquad\quad
 \mu_{bd} = \mu+\tfrac{1}{3}\mu_e-\tfrac{2}{3}\mu_8 \,.
 \end{split}
\end{equation}

At finite temperature and density, the Lorentz invariance is
broken. The three types of four-quark interactions in
Eq.~(\ref{Lagrangian2f}) could develop three different condensates
in a homogeneous phase: $\sigma_\alpha = \langle \bar
q_\alpha^aq_\alpha^a \rangle $ (only sum over $a$), $\Delta =
\langle(\bar{q}_C)_{\alpha}^{a} i \gamma_5 \epsilon^{\alpha \beta
3}\epsilon_{a b 3} q_{\beta}^{b} \rangle$, and $\rho_\alpha =
\langle \bar q_\alpha^a\gamma^{0}q_\alpha^a \rangle $ (only sum
over $a$), where the indies $\alpha$ and  $a$  refer to  flavor
and color, respectively. The last vector-type condensate appears
due to the breaking of Lorentz invariance at finite temperature
and density, which corresponds to the quark number density for
flavor $\alpha$. Here, we follow the common treatment for the
two-flavor CSC phase where the blue quarks do not take part in the
Cooper pairing.

In the mean-field approximation which we will adopt, it is
convenient to introduce the following two gaps and an effective
chemical potential corresponding to the above condensates; that
is, the dynamical quark mass
\begin{equation}
 M(p) =  m - 2G_S (\sigma_u+\sigma_d) g(p) \,,
\end{equation}
the gap for CSC phase
\begin{equation}
\Delta(p) = {\Delta} g(p),
\end{equation}
and the dynamical quark chemical potential
\begin{equation}
{\tilde{\mu}}_\alpha(p) = \mu_\alpha-4G_V{\rho_\alpha} g(p).
\end{equation}
Note that the nonlocal interactions lead to all the gaps being
momentum dependent, which is described by $g(p)$ in the separable
model. The advantage of the choice of the separable interaction is
that the gap equations can be obtained by the variational method.
We also note that the {\em induced  chemical potentials},
$-4G_V{\rho_\alpha} g(p)$, for u
 and d quarks are  different  from each other due to the constraint of electric
charge neutrality ($\mu_d>\mu_u$ and hence $\rho_d>\rho_u$);
notice also, however, that they are dependent only on the
respective  density $\rho_{u,d}$, and hence the dynamical chemical
potentials $\tilde{\mu}_{u, d}(p)$ tend to come closer because
$\rho_d>\rho_u$ with the common coupling constant $G_V$.

Using the standard bosonization technique, the mean-field
thermodynamic potential in the NJL model with the electron
contribution takes the following form:
\begin{equation}
 \Omega =\Omega_L
  -T\sum_n\int\frac{d^3p}{\left(2\pi\right)^3}\mathrm{Tr}\ln
  \frac{{S}_{\text{MF}}^{-1}\left(i\omega_n,\vec{p}\,\right)}{T}+\Omega_{C} \,,
\label{omega}
\end{equation}
where
\begin{equation}
 \Omega_L =2 G_S({\sigma_u^2+\sigma_d^2})-2 G_V({\rho_u^2+\rho_d^2})+\frac{{\Delta^2}}{4G_D}
  -\frac{1}{12\pi^2}\left(\mu_e^4+2\pi^2T^2\mu_e^2
            +\frac{7\pi^4}{15}T^4\right) \,,
\label{omega-L}
\end{equation}
and the sum runs over the Matsubara frequency
 $\omega_n=(2n+1)\pi{T}$ and Tr is taken over color, flavor, and Dirac
 indices. The last term in (\ref{omega-L}) is the contribution of the free electron gas.
Note that a UV divergent counter part $\Omega_{C}$ is introduced
in (\ref{omega}) due to the one-loop integration.

The inverse quark propagator matrix in the Nambu-Gor'kov formalism
is
 given by
 \begin{equation}
  S^{-1}_{\mathrm{MF}}(i\omega_n,\vec{p}) = \bigg(\begin{array}{cc}
  [{G_0^{+}}]^{-1} & \Delta\gamma_5\tau_2\lambda_2 \\
  -\Delta^*\gamma_5\tau_2\lambda_2 &
  [{G_0^{-}}]^{-1} \end{array}\bigg) \,,
\end{equation}
with
\begin{equation}
 [{G_0^{\pm}}]^{-1}=\gamma_0(i\omega_n\pm\hat{\tilde{\mu}}(p))
  -\vec{\gamma}\cdot\vec{p}-\widehat{M}(p) \,.
\end{equation}
Taking the Matsubara sum, we can express the thermodynamic potential
as usual as
\begin{equation}
 \Omega(\mu_e,\mu_3,\mu_8,\sigma,\nu,\Delta;\mu,T)
 =\Omega_L-\sum_{i=1}^{12}\int\frac{d^3p}{(2\pi)^3}\{(E_i-E_i^0)+2T\ln(1+e^{-E_i/T})\},
\label{eqn:therp}
\end{equation}
with the dispersion relations for six quasi-particles (that is, 2
flavors $\times$ 3 colors;  the spin degeneracy is already taken
into account in Eq.~(\ref{eqn:therp})) and six
quasi-anti-particles. In Eq.(\ref{eqn:therp}), the counter part
corresponds to
\begin{equation}
\Omega_C=\sum_{i=1}^{12}\int\frac{d^3p}{(2\pi)^3}E_i^0,
\label{eqn:counter}
\end{equation}
where $E_i^0=E_i(M=m,\Delta=0,\rho=0)$ . The unpaired blue quarks
have the following four energy dispersion relations,
\begin{equation}
 E_{bu} = E - \tilde{\mu}_{bu} \,\quad, \bar{E}_{bu} = E + \tilde{\mu}_{bu} \,\quad
 E_{bd} = E - \tilde{\mu}_{bd} \,\quad, \bar{E}_{bd} = E + \tilde{\mu}_{bd}
\end{equation}
with $E=\sqrt{\vec{p}^2+{M}^2(p)}$.  In the $rd$-$gu$ quark sector
with pairing we can find the four dispersion relations,
\begin{equation}
 \begin{split}
 E_{\text{$rd$-$gu$}}^{\pm} = E_\Delta \pm \tfrac{1}{2}(\tilde{\mu}_{rd}-\tilde{\mu}_{gu})
  = E_\Delta \pm \delta\tilde{\mu}\,,\\
 \bar{E}_{\text{$rd$-$gu$}}^{\pm} = \bar{E}_\Delta \pm
  \tfrac{1}{2}(\tilde{\mu}_{rd}-\tilde{\mu}_{gu})
  = \bar{E}_\Delta \pm \delta\tilde{\mu} \,,
 \end{split}
\end{equation}
and the $ru$-$gd$ sector has another four as follows:
\begin{equation}
 \begin{split}
 E_{\text{$ru$-$gd$}}^{\pm} = E_\Delta \pm \tfrac{1}{2}(\tilde{\mu}_{ru}-\tilde{\mu}_{gd})
  = E_\Delta \mp \delta\tilde{\mu} \,,\\
 \bar{E}_{\text{$ru$-$gd$}}^{\pm} = \bar{E}_\Delta \pm
  \tfrac{1}{2}(\tilde{\mu}_{ru}-\tilde{\mu}_{gd})
  = \bar{E}_\Delta \mp \delta\tilde{\mu} \,,
 \end{split}
\end{equation}
where $E_\Delta=\sqrt{(E-\bar{\tilde{\mu}})^2+\Delta(p)^2}$ and
$\bar{E}_\Delta=\sqrt{(E+\bar{\tilde{\mu}})^2+\Delta(p)^2}$.  The
average chemical potential is defined by
\begin{equation}
 \bar{\tilde{\mu}} = \frac{\tilde{\mu}_{rd}+\tilde{\mu}_{gu}}{2}
 = \frac{\tilde{\mu}_{ru}+\tilde{\mu}_{gd}}{2} = \mu-\frac{\mu_e}{6}
 -2G_V(\rho_u+\rho_d)g(p)+ \frac{\mu_8}{3} \,,
\label{Average}
\end{equation}
and the effective mismatch between the chemical potentials of the
u quark and the d quark takes the form
\begin{equation}
\delta\tilde{\mu}=\tfrac{1}{2}(\mu_e-4G_V(\rho_d-\rho_u)g(p)).
\label{Mismatch}
\end{equation}
We notice that the vector interaction coupled to the difference of
the u and d quark densities tends to diminish the mismatch in the
chemical potentials. This effect will play an important role for
avoiding the color magnetic instability: see below for further
discussions.

For the two-flavor CSC phase, $\mu_3$ always remains zero due to
the left unbroken color SU(2) symmetry for red and green quarks.
Minimizing the thermodynamic potential~(\ref{eqn:therp}), we can
solve the mean fields $\sigma$ , $\Delta$, $\rho_{u[d]}$ together
with the chemical potential $\mu_e$  and $\mu_8$ from
\begin{equation}
 \frac{\partial\Omega}{\partial\sigma}=
 \frac{\partial\Omega}{\partial\Delta_3}=
 \frac{\partial\Omega}{\partial\rho_{u}}=
 \frac{\partial\Omega}{\partial\rho_{d}}=
 \frac{\partial\Omega}{\partial\mu_e}=
 \frac{\partial\Omega}{\partial{\mu_8}}=0\, ,
 \label{gapeq}
\end{equation}
where $\sigma=\sigma_u=\sigma_d$ as demonstrated in
\cite{Zhang:2008wx}. Since $\mu_8$ is very small for color
neutralized two-flavor CSC matter, ignoring this term has a little
effect on the phase structure\footnote{We have checked that the
explicit introduction of $\mu_8$ does not alter the result that
the multiple critical-point structures are realized, although
there is a slight change in the parameter window for realizing
them. It is rather amazing because the value of $\mu_8$ can be in
the same order as that of the induced chemical potential near the
phase boundary. The reason of the robustness of the multiple
critical-point structures lies in the fact that such phase
structures are mainly driven by the competition between the chiral
condensate and the diquark condensate in the coexisting region
while nonzero $\mu_8$ has little effect on this competition.  The
contributions of $\mu_8$ to the red and green quark chemicals have
a negative value, $\mu_8$/3, while that to the blue quark has a
positive value, -$2\mu_8$/3 with a doubled absolute value,
implying  that the effect of $\mu_8$ on the quark mass and diquark
condensate tends to cancel with each other.
 }. Considering this point,
Eq.(\ref{gapeq}) is then simplified as a series of five coupling
equations.

\begin{figure}
\hspace{-.05\textwidth}
\includegraphics*[width=0.5\textwidth]{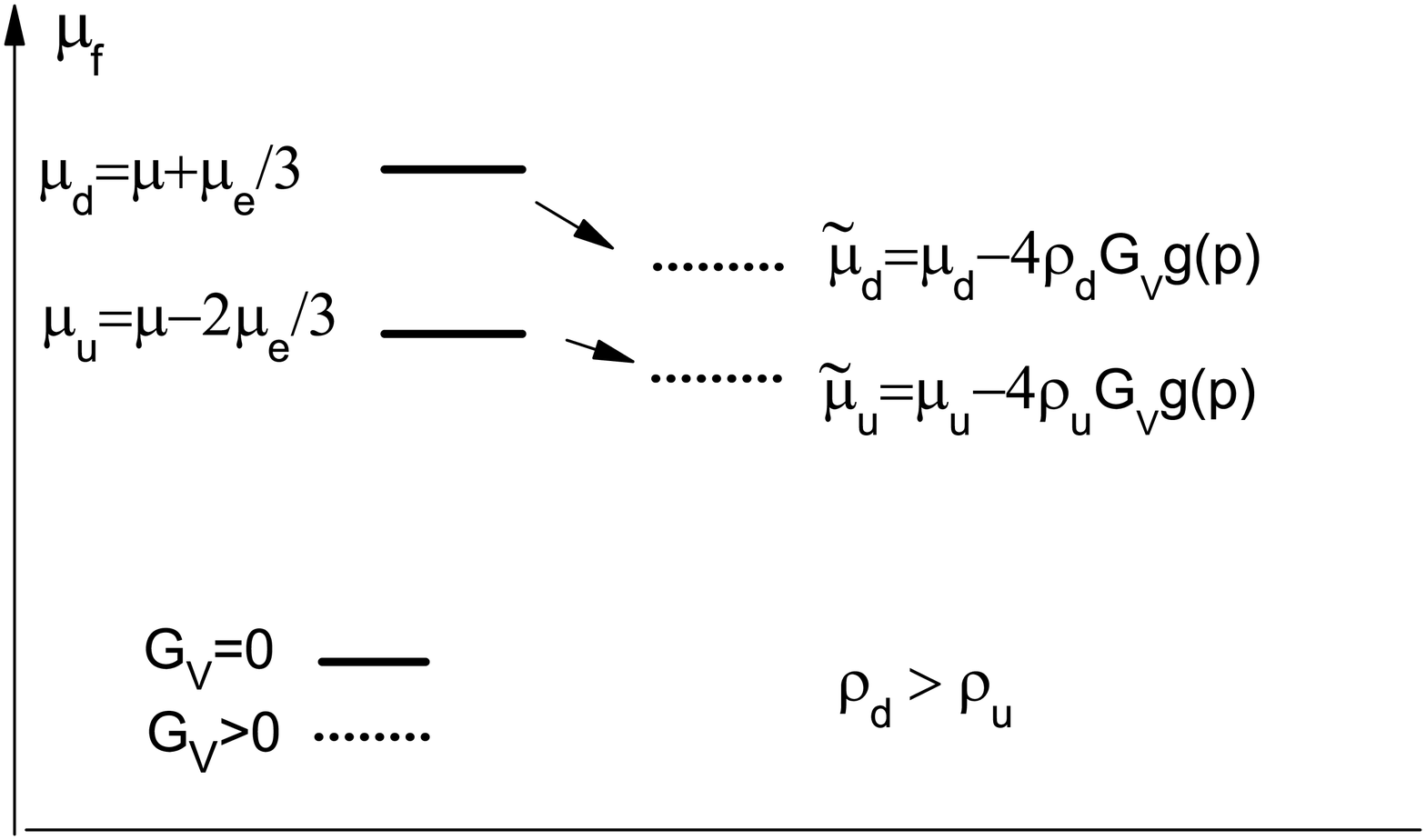}
\caption{Effect of the vector interaction on the
chemical-potential disparity between the u quark and the d quark.
Notice that the amount of the decrease from the chemical potential
to  the effective chemical potential $\tilde{\mu}_{u,d}$ is
proportional to the respective density $\rho_{u,d}$ and hence the
difference between the effective chemical potentials becomes
significantly smaller.} \label{fig:schemetic}
\end{figure}

To determine the region of the unstable homogeneous CSC phase associated
with magnetic instability, one need to calculate the Meissner
masses squared which may be negative with charge neutrality. Here
we adopt the same method as in \cite{Kiriyama:2006jp} to evaluate
the Meissner mass squared
\begin{equation}
m_M^2=\frac{\partial^2}{\partial{B^2}}[\Omega(\Delta)-\Omega(\Delta=0)
]_{B=0}, \label{eqn:Meissner}
\end{equation}
where $B$ has the same meaning as in \cite{Kiriyama:2006jp}. In
our calculation, the upper limit of the  momentum integration in
Eq.(\ref{eqn:Meissner}) is infinity, which helps to  avoid the cut-off
sensitivity for the evaluation of the Meissner masses encountered in the
conventional local NJL model \cite{Kiriyama:2006jp}.

Note that there are three main changes induced by the nonzero
vector-type quark condensates in comparison to the case without
vector interactions \cite{Zhang:2008wx}. First, it give new
negative contributions to the thermal potential, which favors the
phase with relatively larger dynamical quark mass. This effect
becomes more significant when the quark number density is sizable.
Second, it give rise to a negative dynamical chemical potential,
which can delay the chiral restoration towards larger chemical
potential to drive the formation of the coexistence (COE) phase
with both the $\chi\text{SB}$ and the CSC phase. In the COE
region, the competition between the corresponding order parameters
can significantly weaken the first-order chiral phase transition.
Third, as shown in Fig.\ref{fig:schemetic} and briefly mentioned
above,
 the disparity between the densities of u
and d quarks can effectively suppress the chemical-potential
mismatch between these two flavors, which might partially or even
totally cure the chromomagnetic instability. More details of the
influences of the vector interaction combined with the
electric-charge neutrality on the phase of QCD will be given in
the next subsection.


\subsection{ NUMERICAL RESULTS AND DISCUSSIONS} \label{sec:results}

In this subsection, we shall discuss the effect of vector
interaction combined with the charge-neutrality and $\beta$
equilibrium on both the chiral phase transition and the
instability of the CSC phase. Two points will be stressed below:
In general, the model parameter window always exists in the NJL
model, which favors the multiple critical-point structures, and
the number of the critical points can be zero, one, two, three,
and even four; the different dynamic chemical potentials induced
by the vector interaction for u  and d quarks can effectively
shrink the unstable homogeneous CSC region towards  lower
temperatures and larger chemical potentials.

\subsubsection{Multiple critical-point structures for chiral restoration}

In the subsequent subsections we shall present numerical results
and see that what is described above is actually the case.  For
convenience,  we shall adopt the same notations as those in
Refs.~\cite{Hatsuda:2006ps,Zhang:2008wx} to distinguish the
different regions in the $T$-$\mu$ phase diagram: NG, CSC, COE,
and NOR refer to the hadronic (Nambu-Goldstone) phase with
$\sigma\neq0$ and $\Delta=0$, the color-superconducting phase with
$\Delta\neq0$ and $\sigma=0$, the coexisting phase with
$\sigma\neq0$ and $\Delta\neq0$, and the normal phase with
$\sigma=\Delta=0$, respectively, though they have exact meanings
only in the chiral limit.

In Ref.~\cite{Zhang:2008wx}, four types of critical-point structures were
found with varying diquark coupling constant.
We will change the value of the vector coupling $G_V/G_S$
while fixing the diquark coupling at a standard value.
The vector coupling constant in the vacuum may be determined by
the vacuum properties of vector mesons. Since the usually adopted
scale parameter in the two-flavor NJL model is less than the
minimum vacuum mass of vector mesons, one could not get a reliable
vector coupling constant within such formalism; on the other hand,
the vector coupling constant should also be a function of
temperature and quark chemical potential, which is also unknown at
the moment. Therefore, to observe the possible effect of vector
interaction on the phase transition, the vector coupling constant
$G_V/G_S$ is treated as a free parameter rather than a fixed
parameter in the model at hand.

\begin{figure}
\hspace{-.0\textwidth}
\begin{minipage}[t]{.5\textwidth}
\includegraphics*[width=\textwidth]{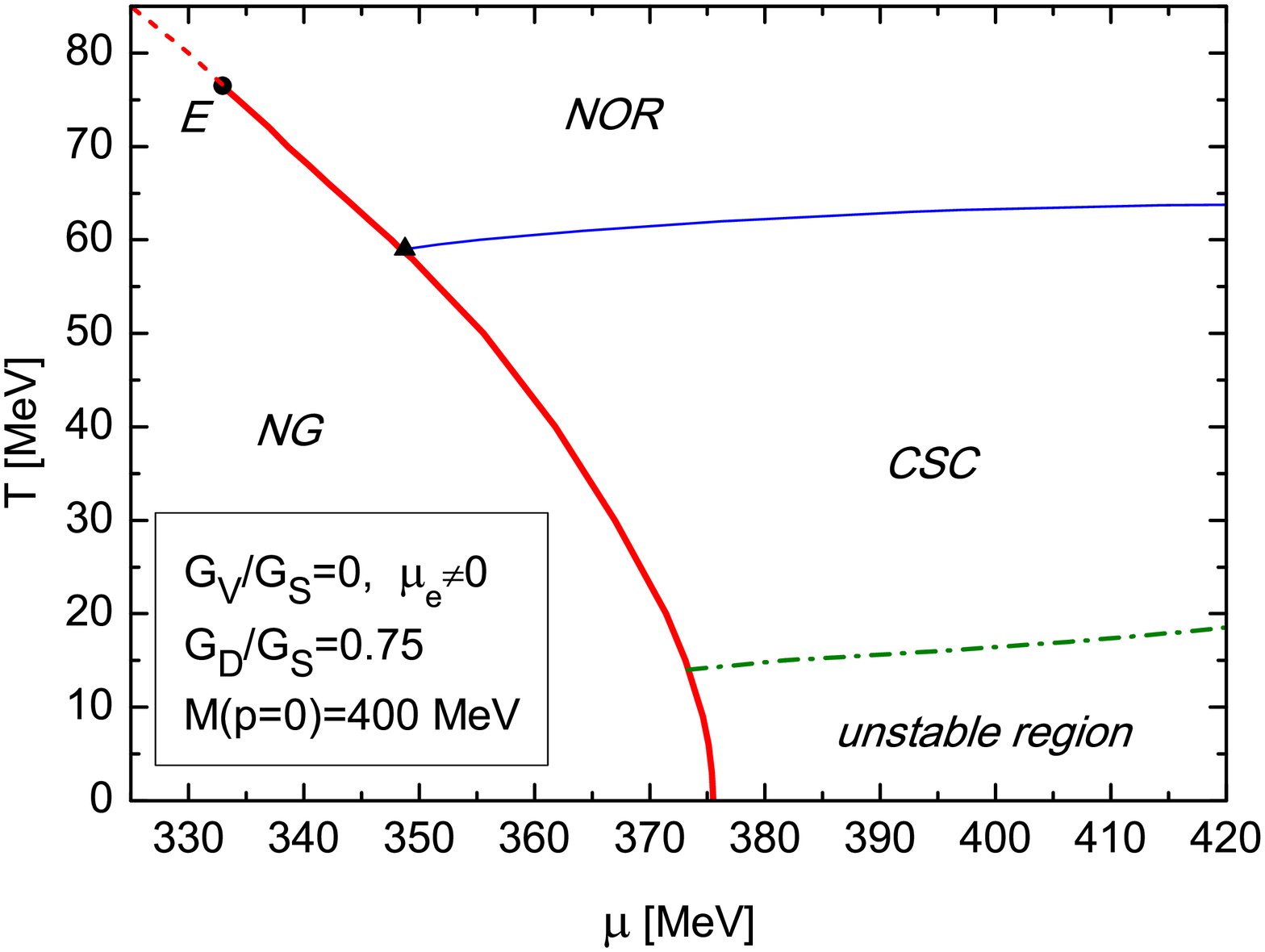}
\centerline{(a)}
\end{minipage}
\hspace{-.05\textwidth}
\begin{minipage}[t]{.5\textwidth}
\includegraphics*[width=\textwidth]{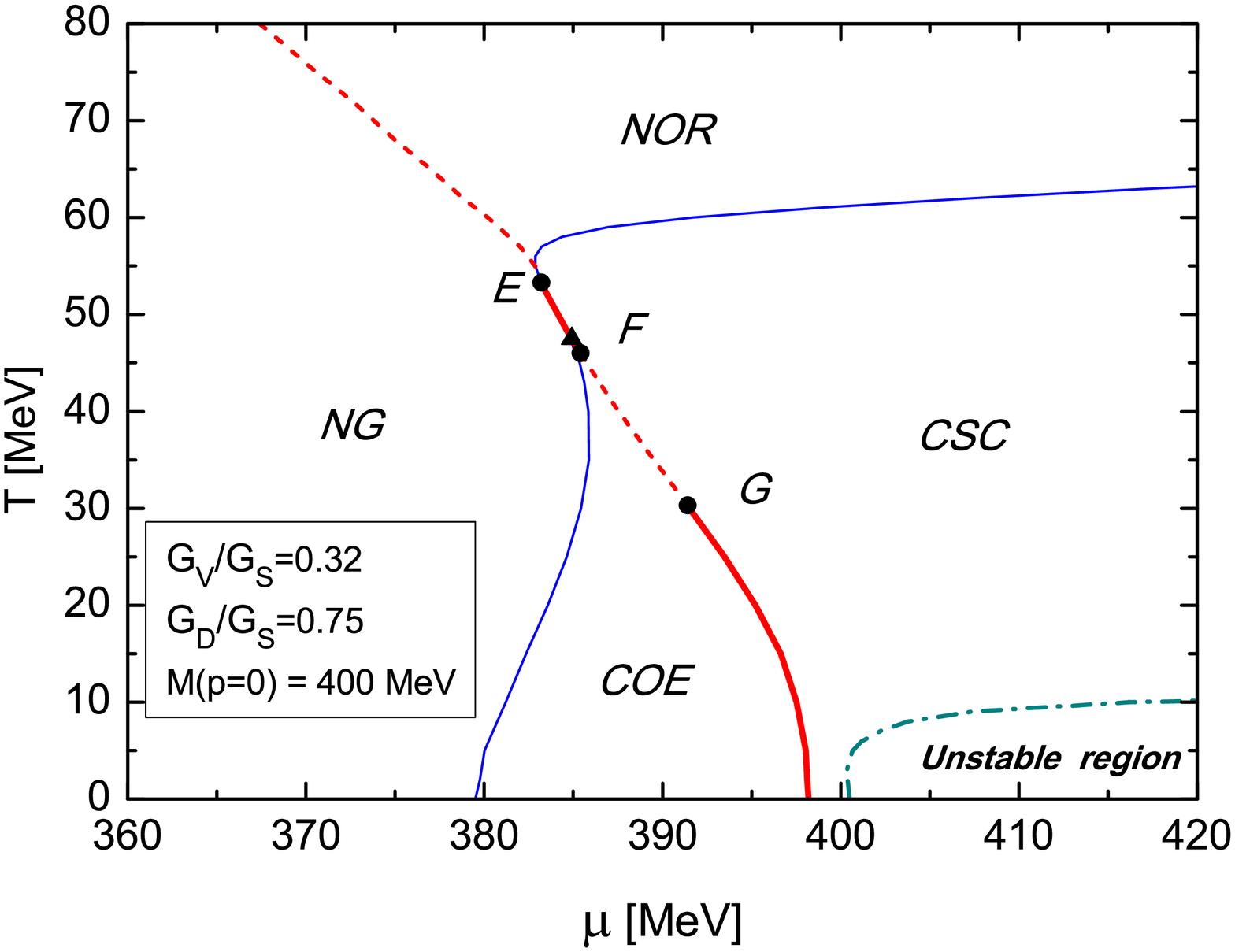}
\centerline{(b)}
\end{minipage}
\hspace{-.1\textwidth}
\begin{minipage}[t]{.5\textwidth}
\includegraphics*[width=\textwidth]{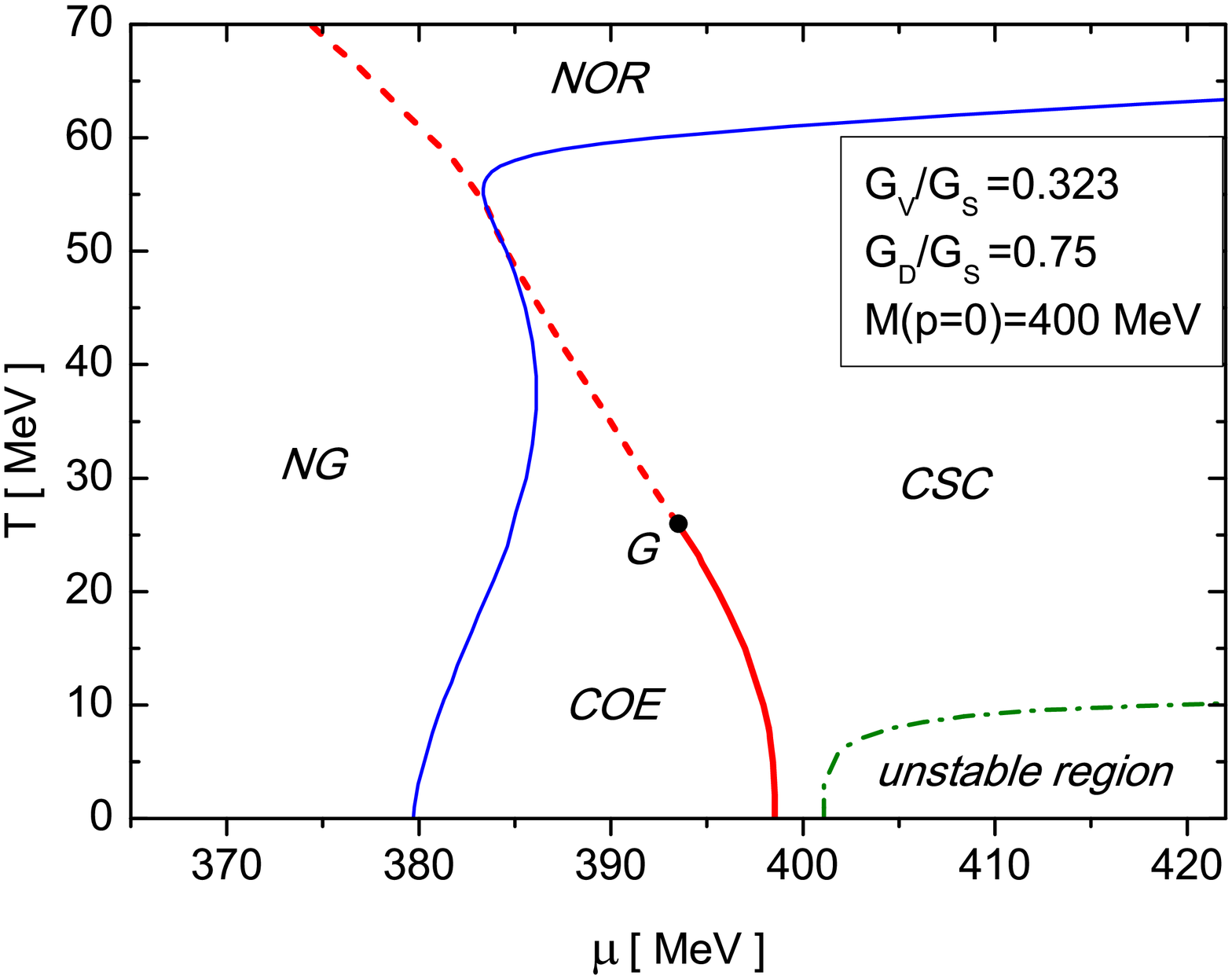}
\centerline{(c)}
\end{minipage}
\hspace{-.05\textwidth}
\begin{minipage}[t]{.5\textwidth}
\includegraphics*[width=\textwidth]{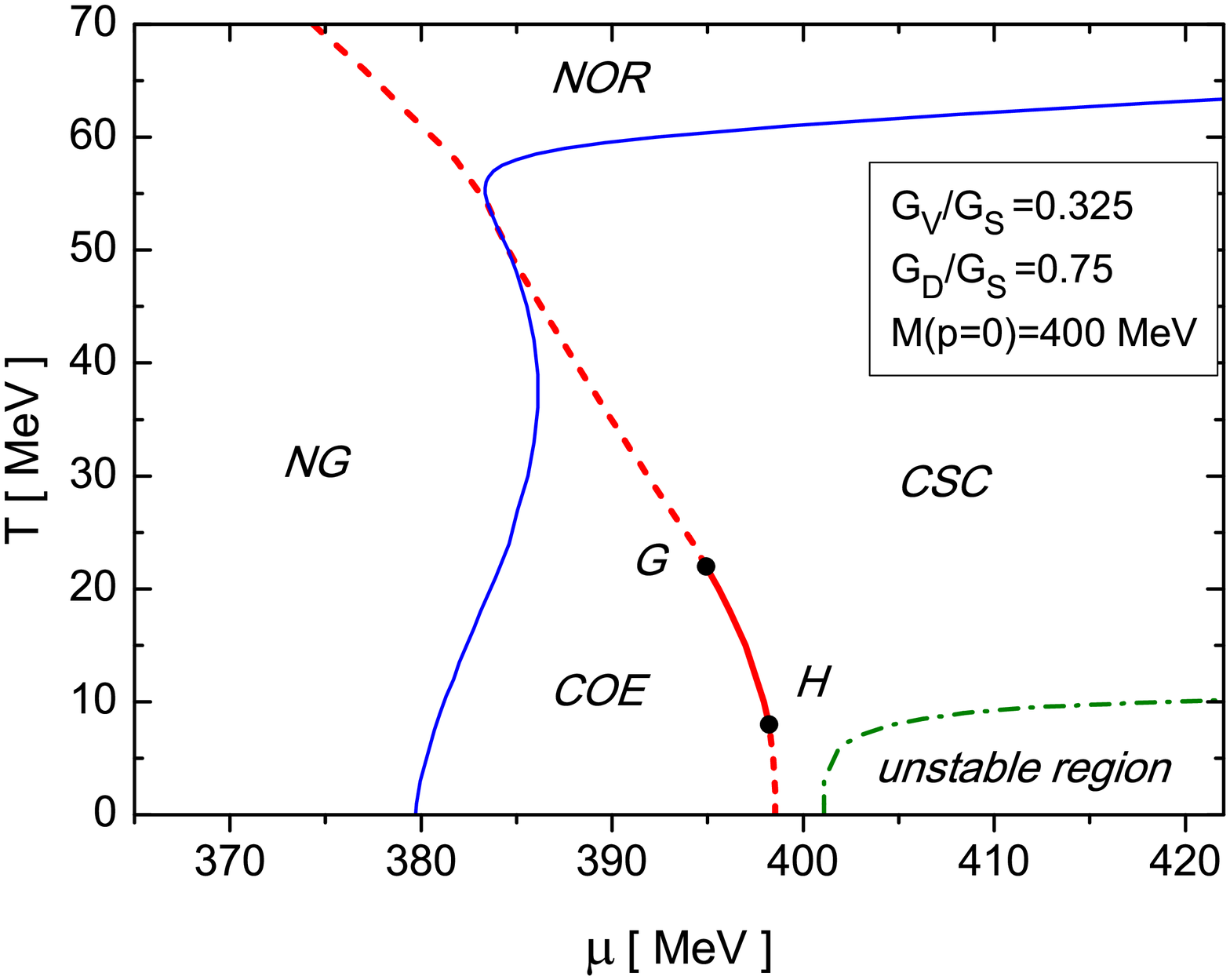}
\centerline{(d)}
\end{minipage}
\caption{The phase diagrams for model parameter set 1 with varying
$G_V/G_S$ and fixed $G_D/G_S=0.75$.
 NG, CSC, COE, and NOR refer to the hadronic (Nambu-Goldstone),
 color-superconducting, coexisting and normal phase, respectively.
 The boundary of the unstable region from
the  chromomagnetic instability is indicated
by the dash-dotted curve. With the increase of
 $G_V/G_S$, the number of critical points changes
 and the instability region tends to shrink toward the lower $T$ and higher
$\mu$ region in the phase diagram.}
\label{fig:pdset1}
\end{figure}

\begin{figure}
\hspace{-.05\textwidth}
\includegraphics*[width=0.65\textwidth]{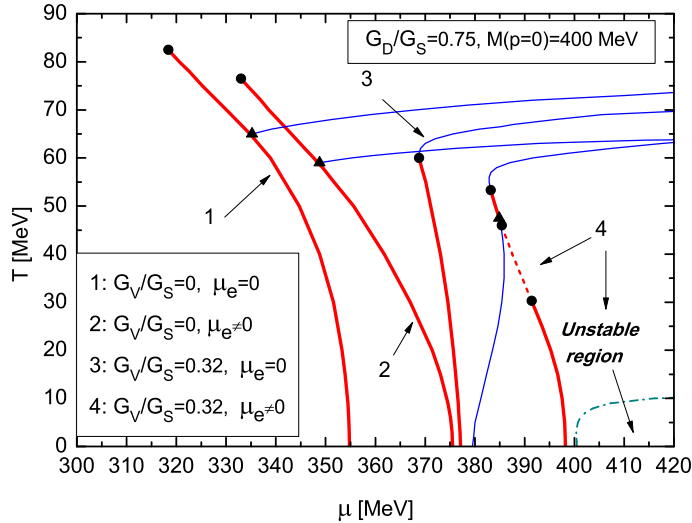}
\caption{Chiral phase transitions for model parameter set 1:
Numbers 1, 2, 3, and 4 correspond to the cases with vanishing
$G_V$ and no charge-neutrality constraint, vanishing $G_V$ and a
charge-neutrality constraint, $G_V/G_S=0.32$ and no
charge-neutrality constraint,  and $G_V/G_S=0.32$ and a
charge-neutrality constraint, respectively. The standard value
$G_D/G_S=0.75$ is adopted in all cases. } \label{fig:set1total}
\end{figure}

The phase diagrams for parameter set 1 in Tabel\ref{tab1} with
different values of $G_V/G_S$  are shown in Fig.\ref{fig:pdset1}.
Parameter set 1 corresponds to  a relatively large  vacuum
constituent quark mass, $M(p=0)=400 \text{MeV}$. Figure
\ref{fig:pdset1}a indicates that chiral restoration at low
temperature  keeps  a first-order transition,  and only a single
chiral critical point appears in the phase diagram  with vanishing
vector interaction. In contrast to the case with the same
$G_D/G_S$ in ~\cite{Zhang:2008wx}, the first-order chiral
transition is too strong to favor the appearance of the COE even
though the finite $\mu_e$ significantly moves the boundary of
chiral restoration towards larger $\mu$ as shown in
Fig.\ref{fig:set1total}. With nonzero $G_V/G_S=0.32$, however,
Fig.~\ref{fig:pdset1}b shows that COE emerges and a
three-critical-point structure for the chiral transition is
realized. This result further verifies the conclusion in
Ref.~\cite{Zhang:2008wx} that the abnormal thermal behavior of the
energy gap for the mismatched diquark pairing can give rise to two
new critical points in the  low temperature region. We notice that
Fig.~\ref{fig:pdset1}b has a similar phase structure to that given
in Fig.4b of ~\cite{Zhang:2008wx},  and the mechanism to realize
this structure is understood much the same way:
 The  first-order transition line E-F is the remnant of
the chiral transition without the CSC phase,  which tends to cease
to exist at high temperature, while the other first-order
transition line ending at G is the chiral transition that survives
the effect of a rather strong diquark condensate at low
temperatures.  In addition,  Fig.\ref{fig:pdset1}b also shows that
all three critical points are free from the chromomagnetic
instability.

Note that the vector interaction with $G_V/G_D=0.32$ is not yet
strong enough to lead to the emergence of COE without the help of
the constraint of electric-charge-neutrality, as indicated in Fig.
\ref{fig:set1total}. In contrast to the case without both $G_V$
and the constraint of electric charge, Fig. \ref{fig:set1total}
shows that the critical chemical potential for the chiral
transition at zero temperature is delayed towards larger $\mu$  by
about $42\text{MeV}$ by nonzero $G_V$ and $\mu_e$. Therefore, the
combined influence of these two elements on the chiral transition
is quite significant.

By further increasing $G_V$, Fig.\ref{fig:pdset1}d shows that the
remnant first-order chiral transition turns into a crossover and
the surviving first-order transition  still  exists with the
emergence of another new end point,  H. This is quite different
from the result in Ref.\cite{Zhang:2008wx} that the surviving part
of the first boundary initially vanishes with increasing $G_D$.
The reason for the difference is that the relatively large vector
interaction has more significant impact on the location of the
critical point E, which shifts to lower $T$ and larger $\mu$ and
eventually is eaten by the enlarged COE. This point is also
indicated clearly in Fig.\ref{fig:set1total}. The appearance of H
is due to the stronger competition between the chiral condensate
and the diquark condensate at lower temperatures. Note that the
two critical points in Fig.\ref{fig:pdset1}d are also free from
the chromomagnetic instability. Figure.\ref{fig:pdset1}c indicates
that one critical-point structure appears again in a very narrow
parameter region of the vector coupling, where the first-order
line only contains the surviving part.

For  larger vector coupling, the chiral boundary will totally
become crossover, which is not shown in Fig.\ref{fig:pdset1}. In
conclusion, with parameter set 1 and $G_D/G_S=0.75$ , five types
of  critical-point structures with critical-point numbers 1, 3, 1,
2,  and 0 are found in the model
when $G_V/G_S$ is varied in the range $0-0.5$. In contrast to the
situation in Ref.\cite{Zhang:2008wx}, all the multiple critical
points obtained with parameter set 1 are far from the unstable
homogeneous CSC region.

\begin{figure}
\hspace{-.0\textwidth}
\begin{minipage}[t]{.5\textwidth}
\includegraphics*[width=\textwidth]{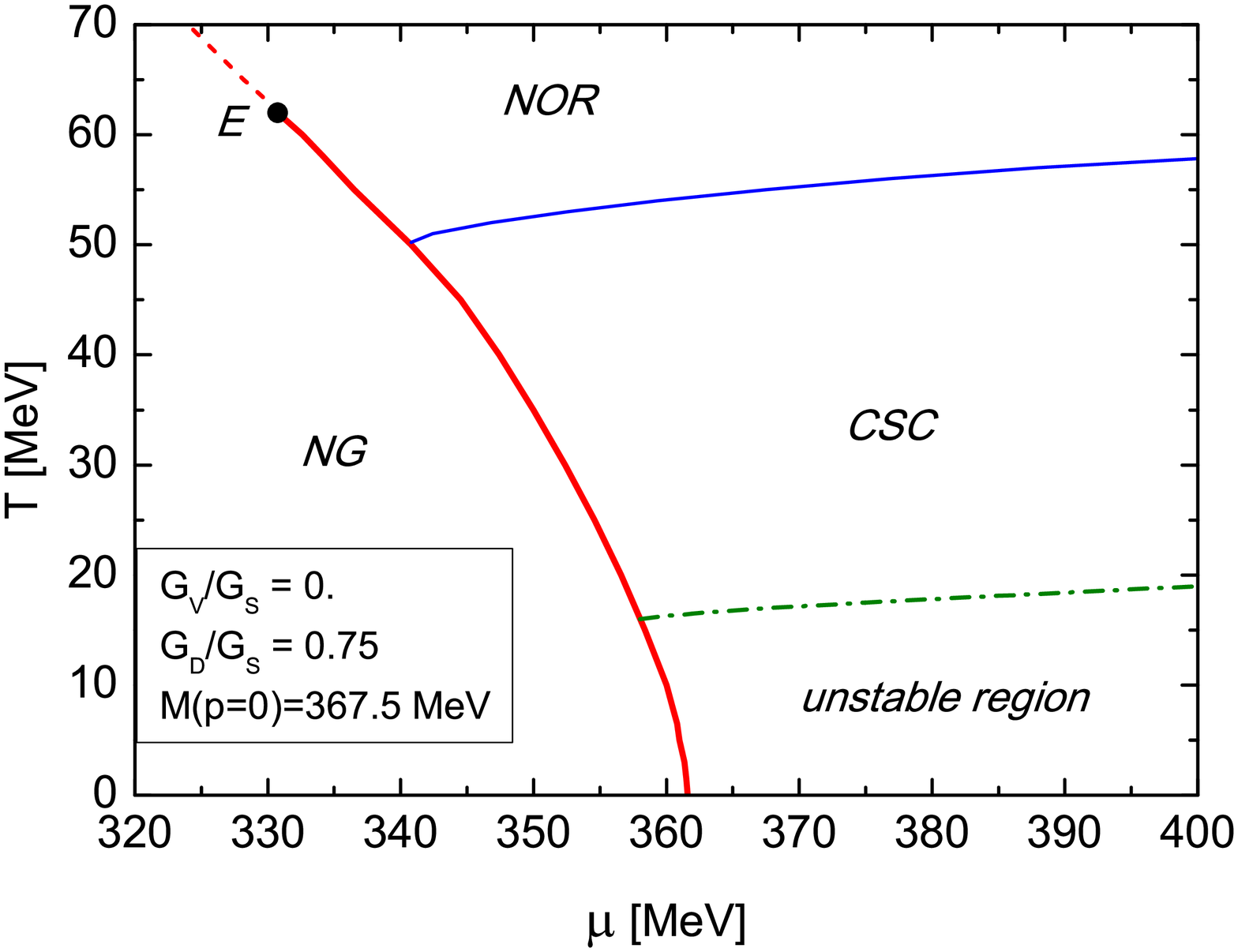}
\centerline{(a)}
\end{minipage}
\hspace{-.05\textwidth}
\begin{minipage}[t]{.5\textwidth}
\includegraphics*[width=\textwidth]{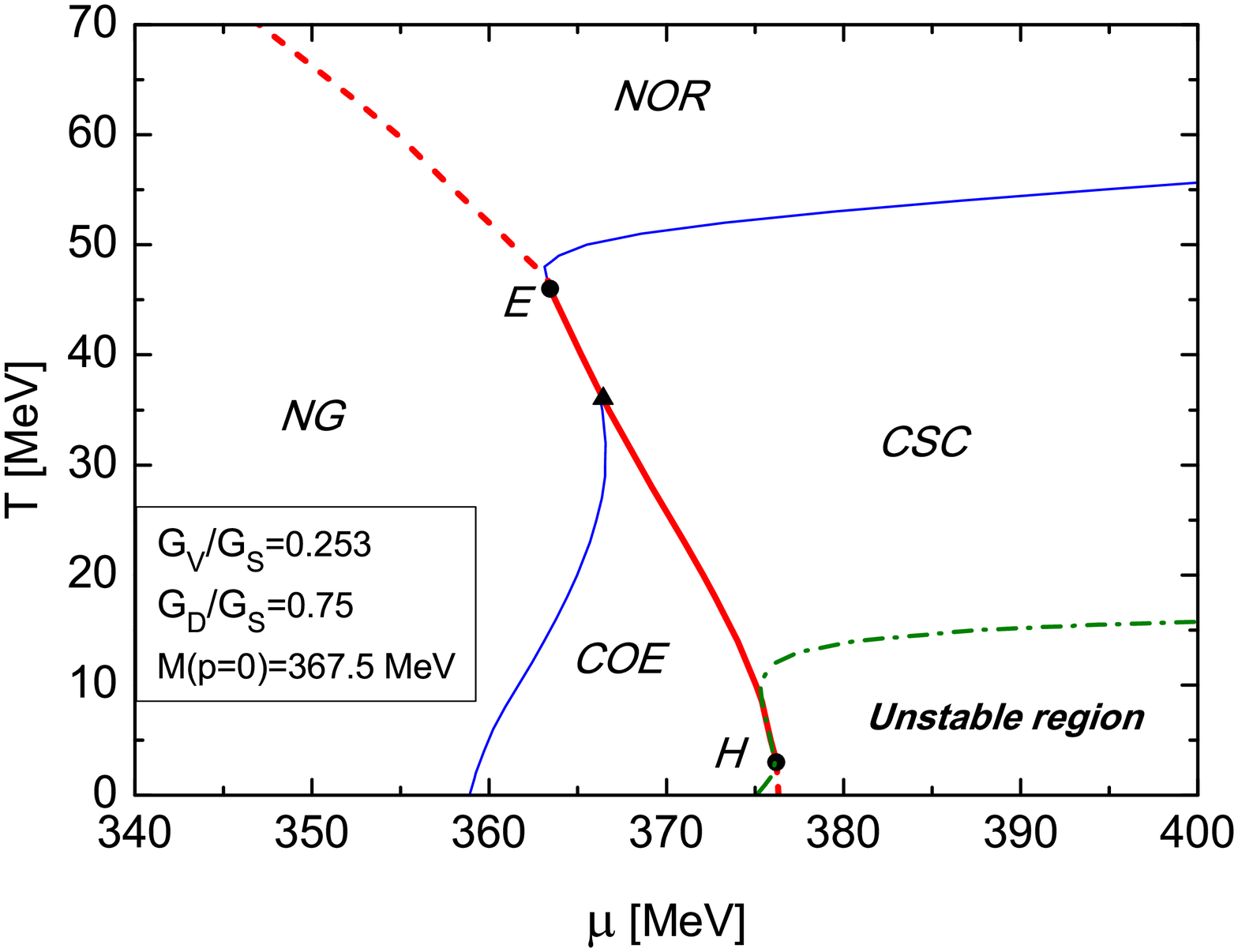}
\centerline{(b)}
\end{minipage}
\hspace{-.1\textwidth}
\begin{minipage}[t]{.5\textwidth}
\includegraphics*[width=\textwidth]{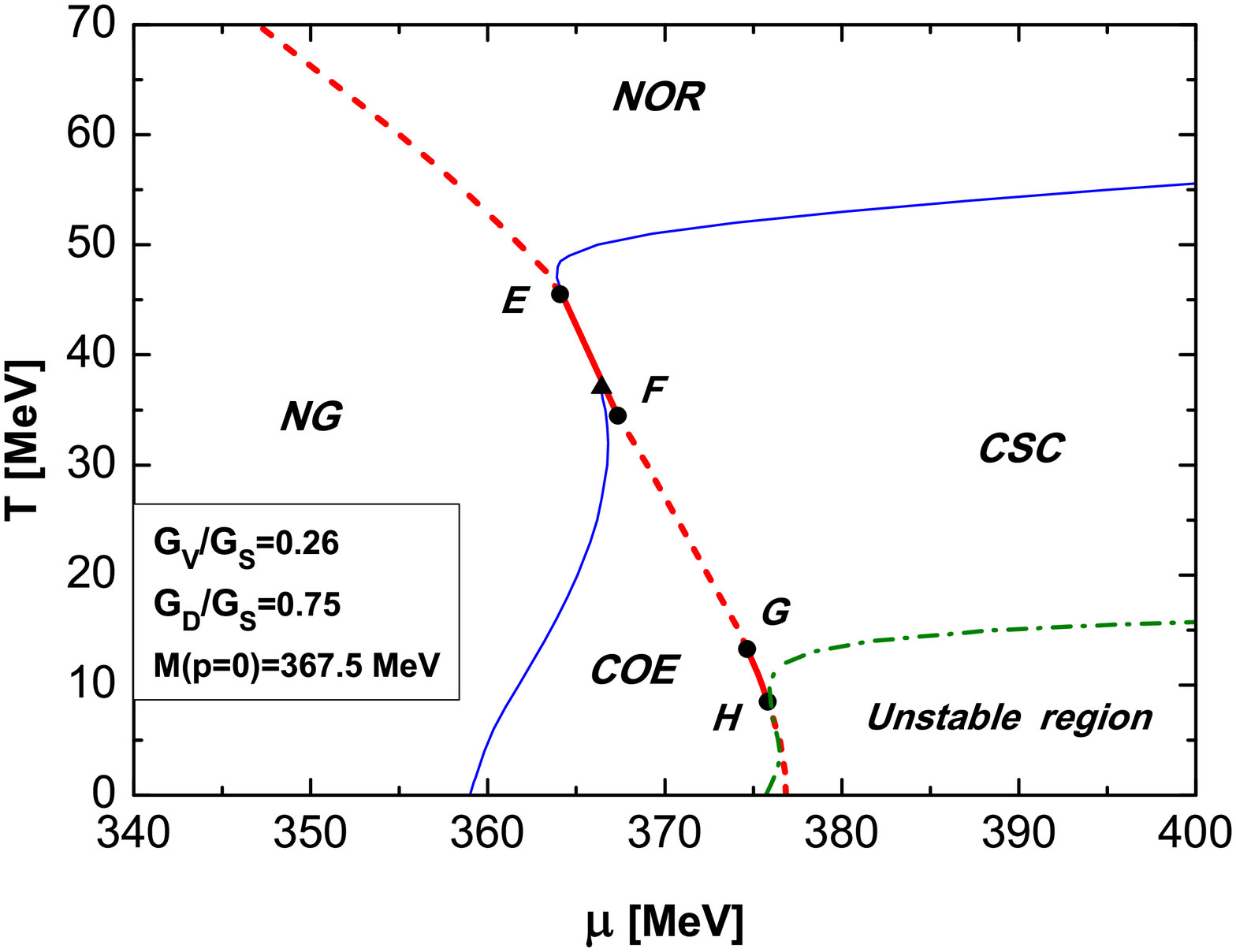}
\centerline{(c)}
\end{minipage}
\hspace{-.05\textwidth}
\begin{minipage}[t]{.5\textwidth}
\includegraphics*[width=\textwidth]{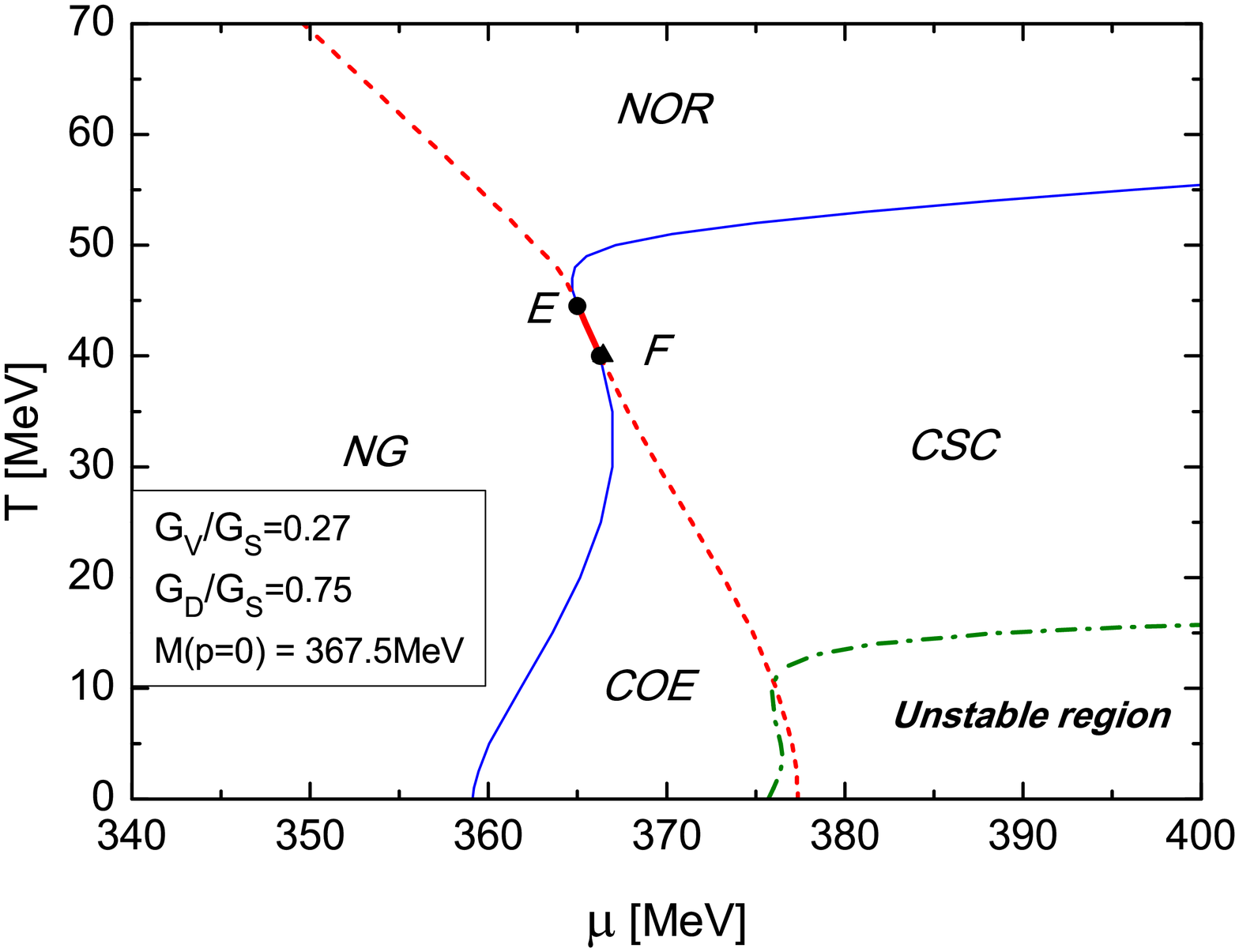}
\centerline{(d)}
\end{minipage}
\caption{The phase diagrams for model parameter set 2 with varying
$G_V/G_S$ and fixed $G_D/G_S=0.75$.  The unstable region with
chromomagnetic instability is indicated by the dash-dotted curve.}
\label{fig:pdset2}
\end{figure}

The phase diagrams for parameter set 2 with varying $G_V/G_S$ are
shown in Fig.\ref{fig:pdset2}. Compared to parameter set 1, the
resulting vacuum quark mass $M(p=0)$ is reduced and the
first-order chiral transition is not very strong. For vanishing
$G_V$, Fig.\ref{fig:pdset2}a gives a similar phase diagram as
Fig.\ref{fig:pdset1}a. When $G_V/G_S$ is increased to 0.253,
Fig.\ref{fig:pdset1}b tells us that  a new critical point, H,
appears  at a very low temperature. This is not surprising since
the larger the diquark condensate in the COE phase, the more
suppressed the chiral condensate in the lower $T$ and larger $\mu$
region, where the phase change is a crossover. With the further
increased vector coupling, Fig.\ref{fig:pdset2}c shows that {\em
four} critical points appear on the chiral boundary: Owing to the
abnormal $T$ dependence of the diquark condensate, two new
critical points denoted by F and G exist in the phase diagram.
Because the vector interaction is not as strong as the case in
Fig.\ref{fig:pdset1}c,
both the remnant and the surviving first-order chiral transition
remain in the phase diagram. One can see from
Fig.\ref{fig:pdset2}b and Fig.\ref{fig:pdset2}c that the critical
point labeled as H is located on the border between the stable
region and the unstable region, while other critical points are
free from the chromomagnetic instability. In
Fig.\ref{fig:pdset2}d,  with much lager $G_V/G_S$, the surviving
first-order transition in the lower $T$ region can not survive
anymore,  and only the remnant first-order transition in the high
$T$ region is left on the chiral boundary. Therefore, for
parameter set 2 and the standard diquark coupling, five types of
chiral critical point structures  also exist,  and the order of
the number of the critical points is 1,2,4,2,0 with increasing
vector interaction.

\begin{figure}
\hspace{-.0\textwidth}
\begin{minipage}[t]{.5\textwidth}
\includegraphics*[width=\textwidth]{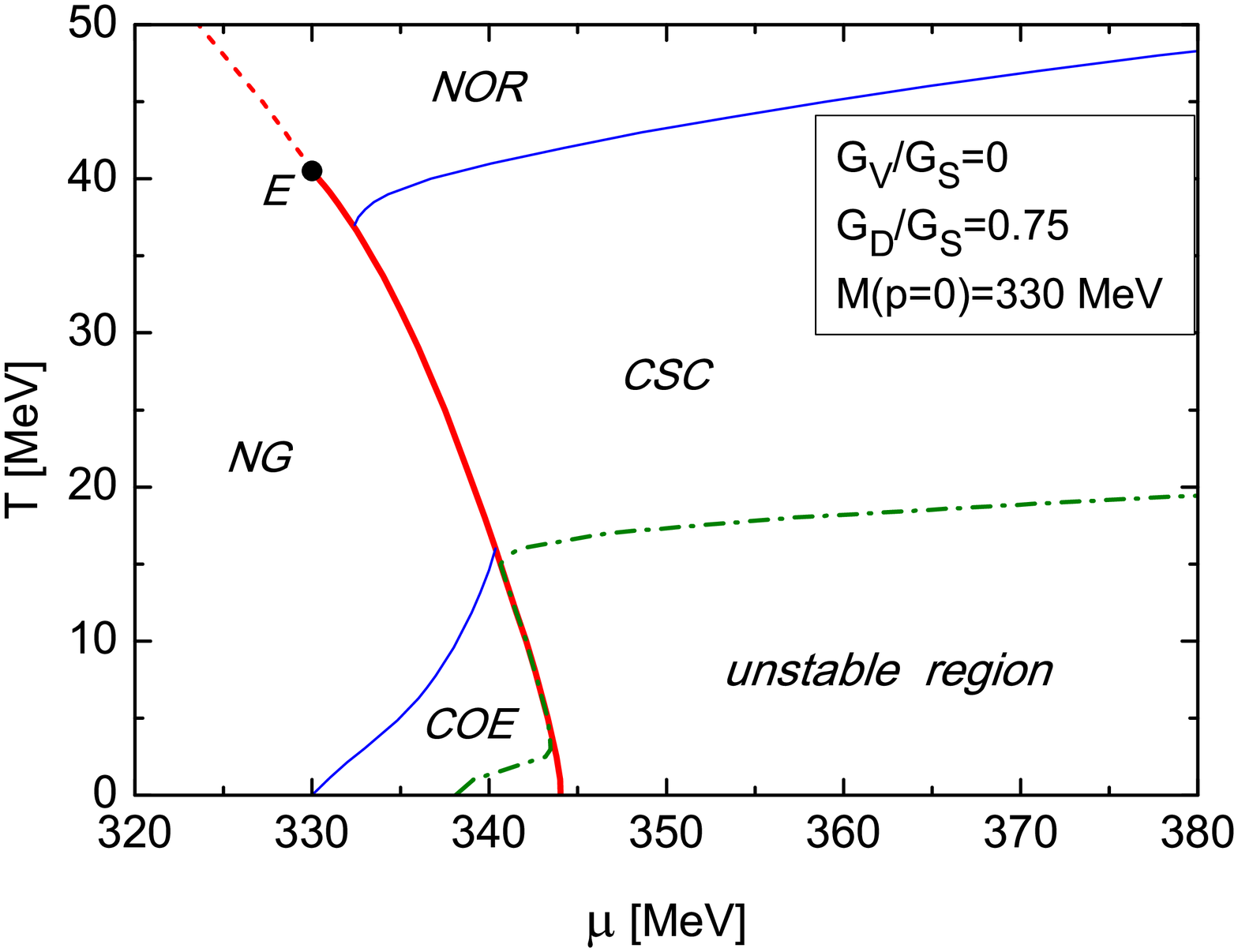}
\centerline{(a)}
\end{minipage}
\hspace{-.05\textwidth}
\begin{minipage}[t]{.5\textwidth}
\includegraphics*[width=\textwidth]{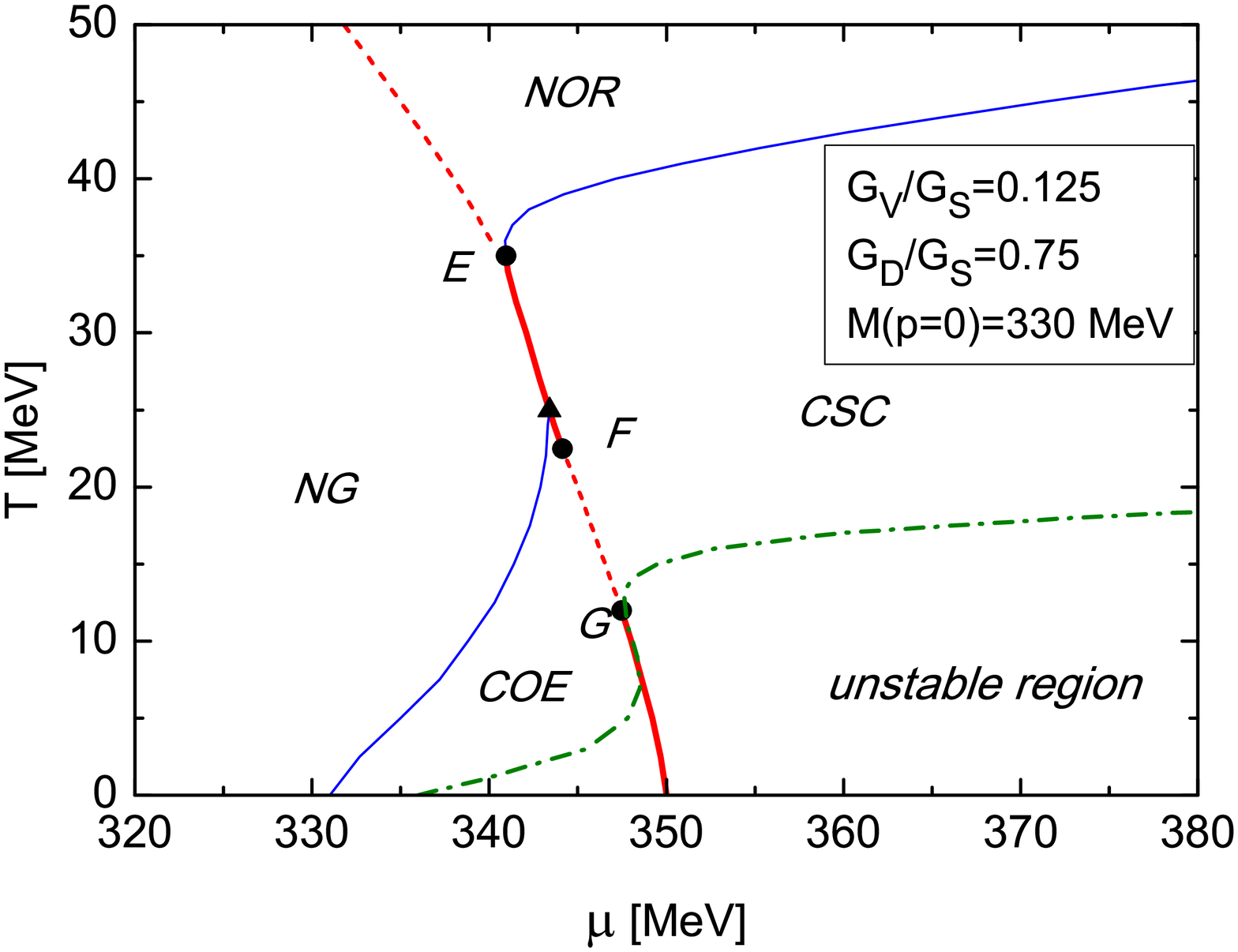}
\centerline{(b)}
\end{minipage}
\hspace{.0\textwidth}
\begin{minipage}[t]{.5\textwidth}
\includegraphics*[width=\textwidth]{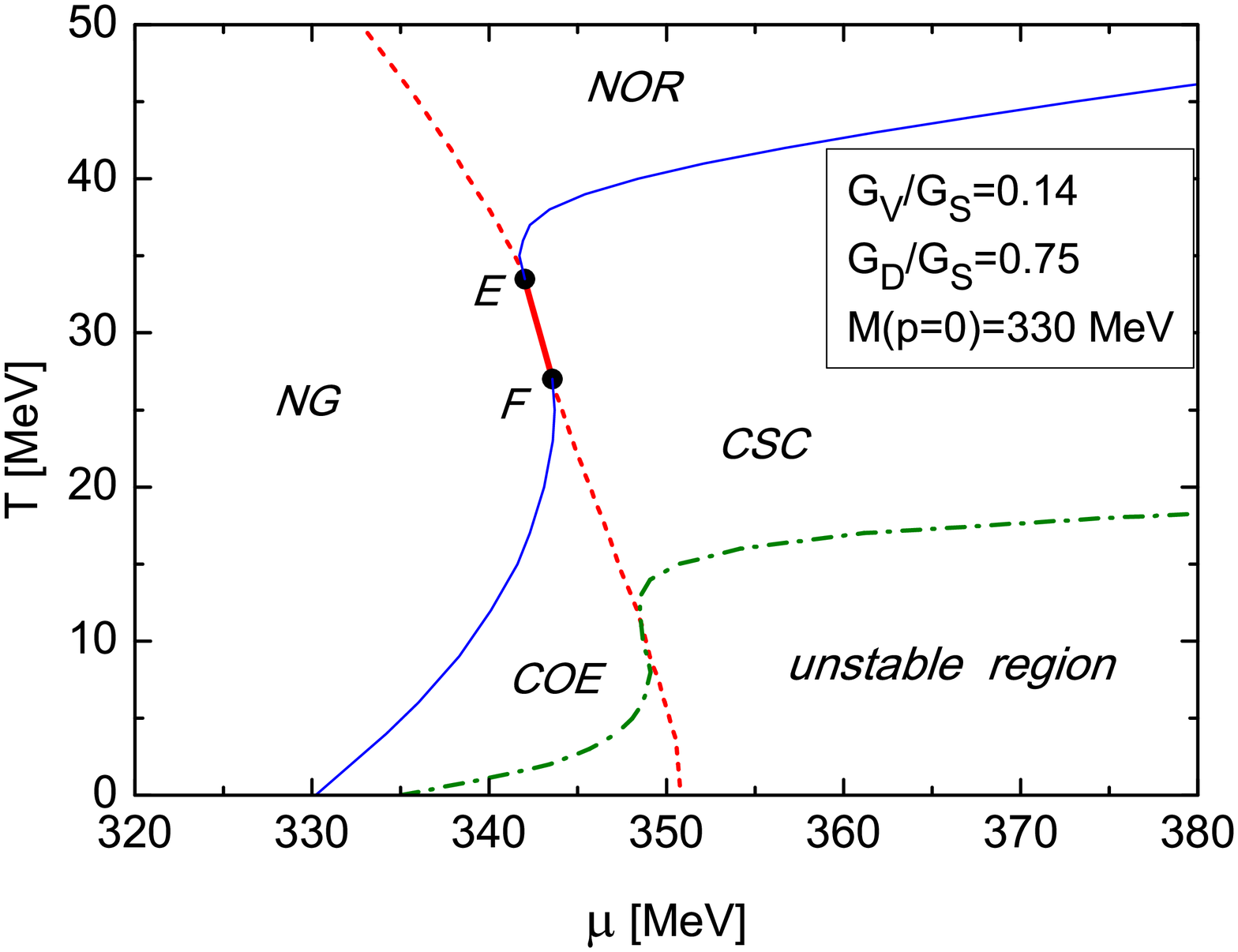}
\centerline{(c)}
\end{minipage}
\caption{The phase diagrams for model parameter set 3 with varying
$G_V/G_S$ and fixed $G_D/G_S=0.75$.  The unstable region with
chromomagnetic instability is indicated by the dash-dotted curve.}
\label{fig:pdset3}
\end{figure}

In Table \ref{sec: Two flavor }, model parameter set 3 reproduces
the smallest vacuum dynamical quark mass,  and the corresponding
first-order chiral transition is also the weakest one.
Figure.\ref{fig:pdset3}a shows that only finite $\mu_e$ is capable
of realizing the COE in the phase diagram, even though the
competition between the chiral and CSC correlations in the COE is
not strong enough to lead to a multiple critical-point structure.
With the help of vector interaction, the three- and
two-critical-point structures appear in Figs. \ref{fig:pdset3}b
and \ref{fig:pdset3}c, respectively. In contrast to the former two
cases, only a very small vector interaction can lead to multiple
critical-point structures.

Since all the critical-point structures mentioned above are
obtained within the range $G_V/G_S<0.35$, one can expect that
similar results may be obtained for a relatively weak diquark
coupling with the vector coupling varying in the range
$0<G_V/G_S<0.5$. In general, at least in the two-flavor NJL model,
we can conclude that there always exists a region in the
parameters plane of $G_D$-$G_V$ which favors a multiple
chiral-critical-point structure,  with color superconductivity and
electric neutrality being taken into account.

Here it is worth emphasizing that the abnormal $T$ dependence of
the diquark condensate is a general feature in the COE region,
even though the same behavior only happens in the CSC dominant
region with a weak diquark coupling. The reason is that the quark
Fermi surface in the COE is relatively small due to the sizable
quark mass. Note that, the various critical-point structures
demonstrated above are realized in a very simple chiral quark
model of QCD with the physical constraint of charge neutrality.
Because of the complexity of QCD, it is possible that there are
more than one chiral critical point in the true QCD phase diagram
.

\subsubsection{Suppressing the chromomagnetic instability  }

It is well known that the asymmetric homogeneous g2CSC phase
suffers from the chromomagnetic instability. At zero temperature,
the calculation based on the hard-dense-loop
method~\cite{Huang:2004bg} suggests that the  Meissner mass
squared of the 8th gluon becomes negative for
$\frac{\delta\mu}{\Delta}>1$ while the 4th-7th gluons acquire
negative Meissner masses squared for
$\frac{\delta\mu}{\Delta}>1/\sqrt{2}$. Note that without the
vector interaction, the chemical-potential mismatch $\delta\mu$ is
just equal to $\mu_e/2$.
The instability of the homogeneous CSC phase should imply the
existence of a yet unknown but stable phases  in this region of
the phase diagram.
Such  examples  proposed so far include
 the Larkin-Ovchinnikov-Fulde-Ferrel (LOFF)
phase \cite{Giannakis:2004pf} and the gluonic phase
\cite{Gorbar:2005rx}.

Because the Fermi surface is smeared by finite temperature, the
 homogeneous neutral two-flavor CSC phase can be stable in
a somewhat higher temperature region
\cite{Kiriyama:2006jp,Kiriyama:2006ui,He:2007cn}. The same thing
happens for the homogeneous neutral CFL phase with three flavors
\cite{Fukushima:2005cm}. It is also reported in
Ref.\cite{Kitazawa:2006zp} that a large quark mass and a strong
coupling can effectively suppress the instability even at zero
temperature. Here we will demonstrate that the repulsive vector
interaction can also resolve or suppress the instability problem.
The reason is very simple: According to Eq.~(\ref{Mismatch}), the
density mismatch between u and d quarks gives a negative
contribution to $\delta\mu$, which can effectively suppress the
ratio $\delta\mu/\Delta$.

\begin{figure}
\hspace{-.0\textwidth}
\begin{minipage}[t]{.5\textwidth}
\includegraphics*[width=\textwidth]{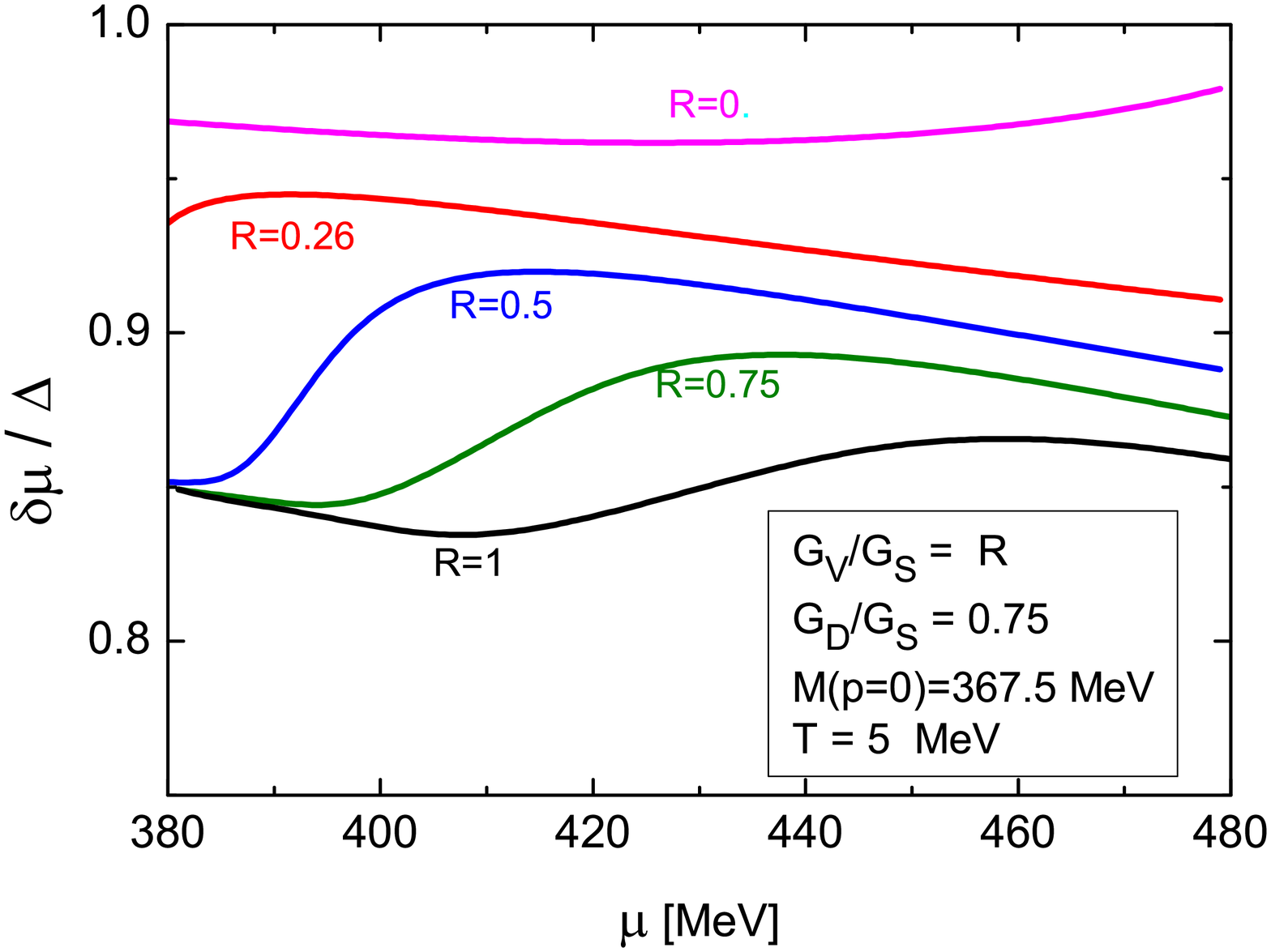}
\centerline{(a)}
\end{minipage}
\hspace{-.05\textwidth}
\begin{minipage}[t]{.5\textwidth}
\includegraphics*[width=\textwidth]{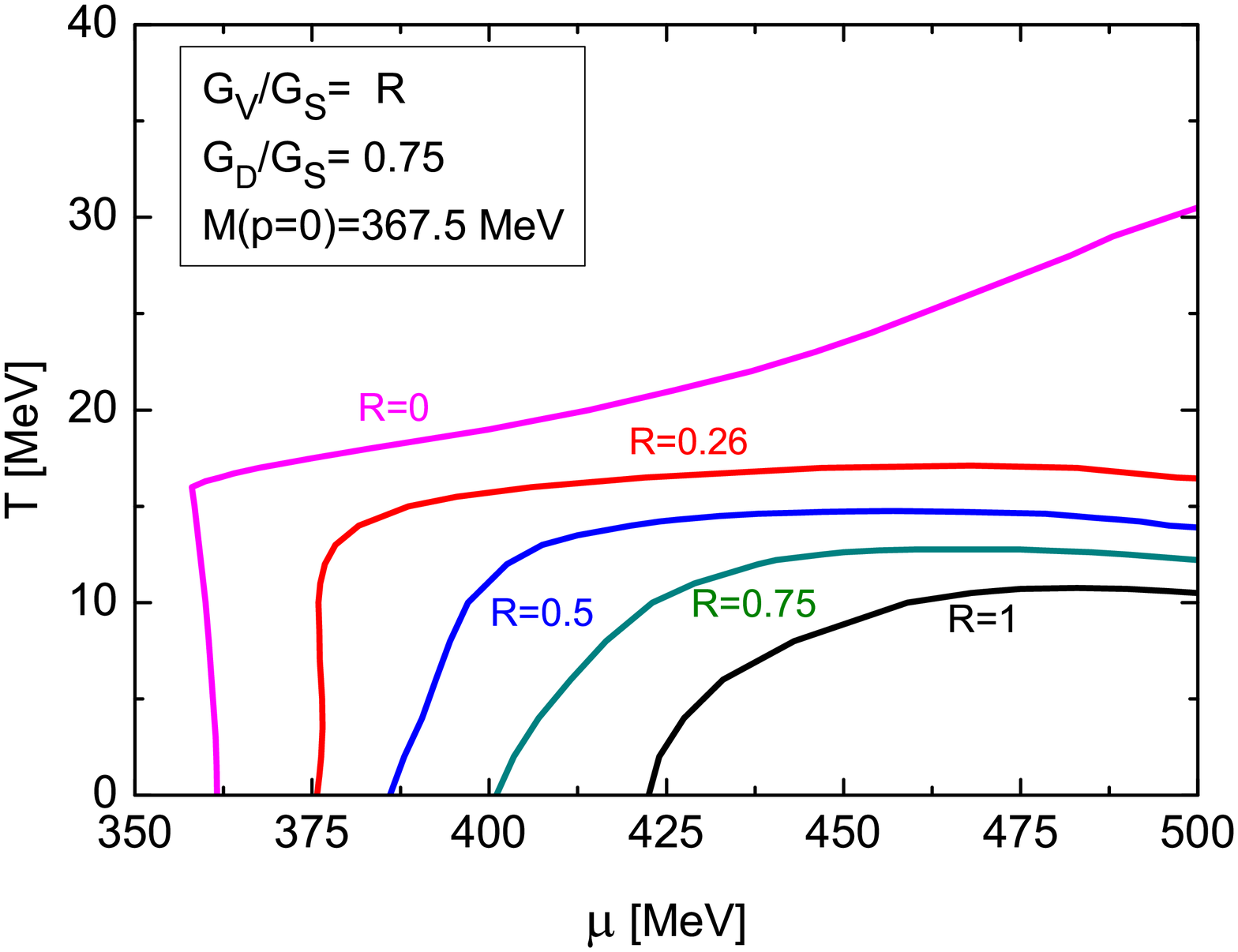}
\centerline{(b)}
\end{minipage}
\caption{The effect of the repulsive vector interaction on the
ratio $\delta\mu/\Delta$ (left figure) and the location of the
boundary between the stable and unstable homogeneous regions(right
figure). With the increase of the ratio $G_V/G_s\equiv R$,
 $\delta\mu/\Delta$ tends to become smaller and the unstable
region in the $T$-$\mu$ plane shrinks toward a lower $T$ and
higher $\mu$ region.} \label{fig:unstable-stable}
\end{figure}

The ratio $\delta\mu/\Delta$ (here the ratio
${\delta\mu}/{\Delta}$ refers to ${\delta\mu(p=0)}/{\Delta(p=0)}$)
as a function of $\mu$ with fixed $T=5\text{MeV}$ for different
$G_V/G_S$ is shown in Fig.\ref{fig:unstable-stable}a, where model
parameter set 2 is used with $G_D/G_S=0.75$. One can see that
$\delta\mu/\Delta$ significantly decreases with an increasing
vector coupling. The regions of  the chromomagnetic instability
for the corresponding $G_V/G_S$ are shown in
Fig.\ref{fig:unstable-stable}b. With increasing $G_V/G_S$, the
unstable region shrinks towards a lower-temperature and higher
chemical-potential region. Actually, we have already seen that the
same phenomenon occurs in the phase diagrams of
Figs.\ref{fig:pdset1}-\ref{fig:pdset3}.

We should notice here that,  as seen from  Eq.(\ref{di}), the
repulsive vector interaction also suppresses the magnitude of the
diquark condensate due to the reduced effective quark chemical
potential. However, the direct effect of the vector interaction on
$\delta\mu$ is more significant than that on $\Delta$, in
particular for finite temperature. We stress that this important
role of the vector interaction in the CSC phase, especially for
the instability problem, was first revealed in this work. Of
course, the vector interaction may not totally remove the unstable
region from the phase diagram unless the  diquark and/or vector
coupling are very large. Therefore,  other mechanism may still be
necessary for a thorough cure of the magnetic instability.

\section{Two-plus-one-flavor case}
\label{sec: Two-plus-one } In this section, the study of the
influence of the vector interaction on the chiral phase transition
is extended to the two-plus-one-flavor NJL formalism.

\begin{figure}
\hspace{-.0\textwidth}
\begin{minipage}[t]{.5\textwidth}
\includegraphics*[width=\textwidth]{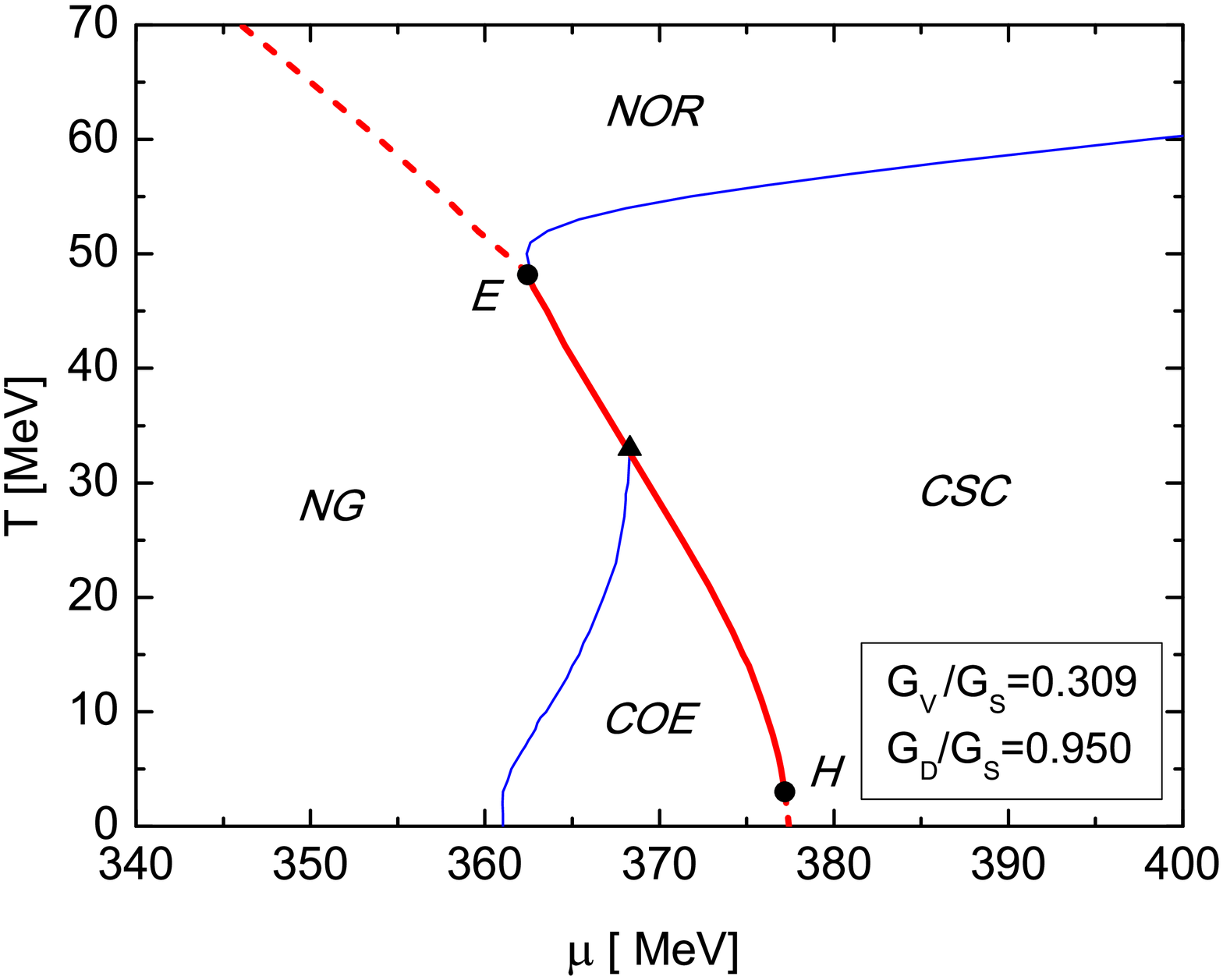}
\centerline{(a)}
\end{minipage}
\hspace{-.05\textwidth}
\begin{minipage}[t]{.5\textwidth}
\includegraphics*[width=\textwidth]{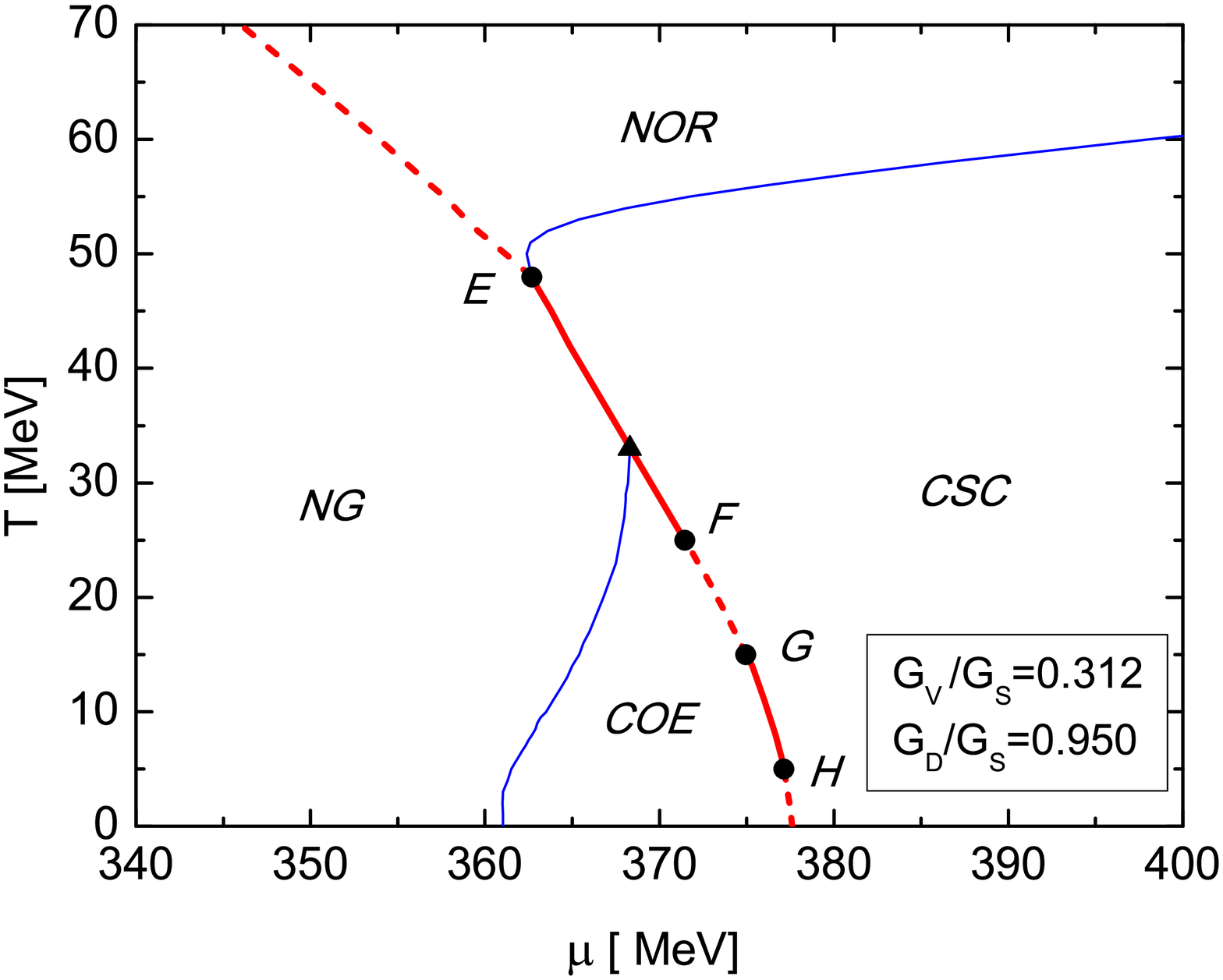}
\centerline{(b)}
\end{minipage}
\hspace{-.0\textwidth}
\begin{minipage}[t]{.5\textwidth}
\includegraphics*[width=\textwidth]{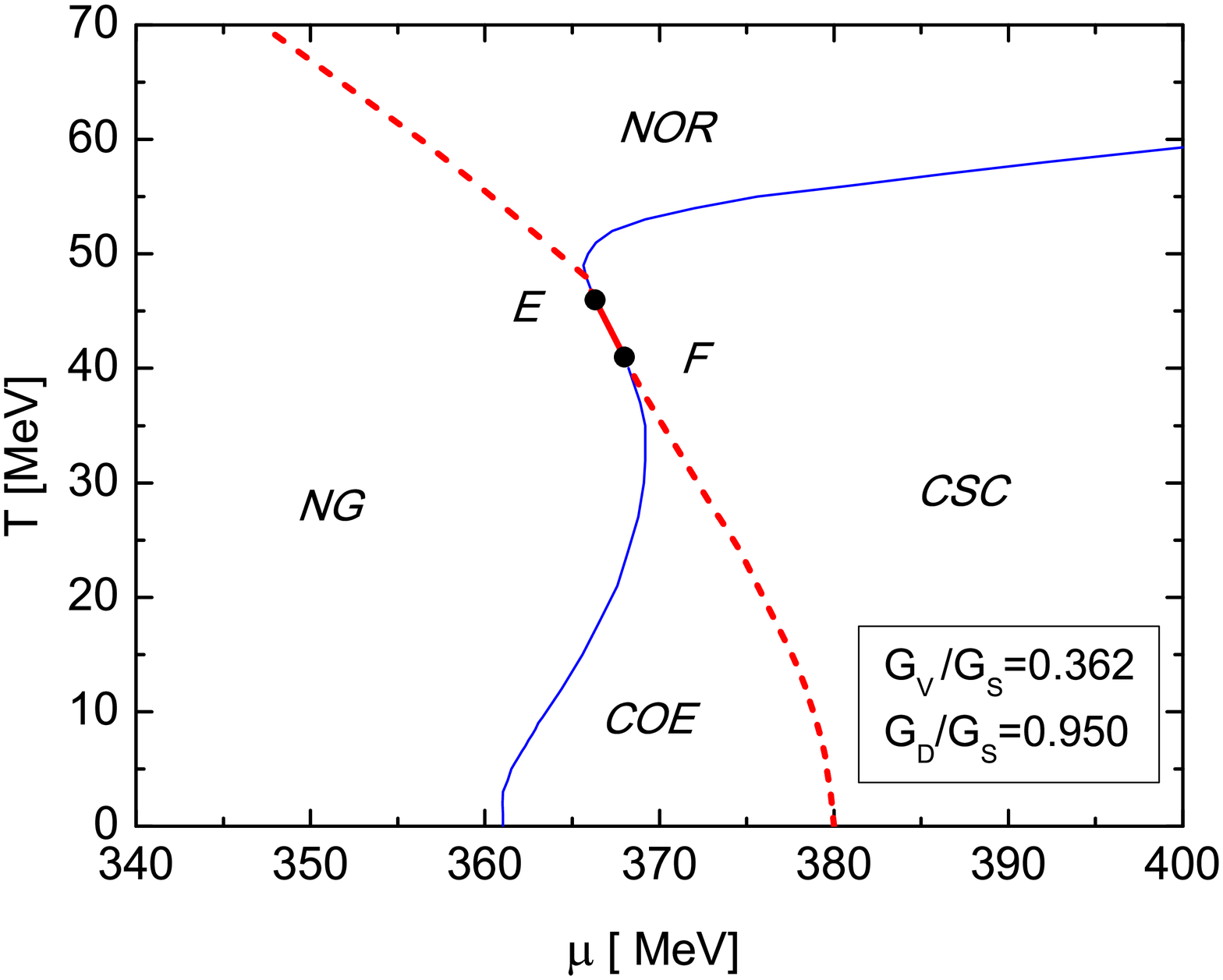}
\centerline{(c)}
\end{minipage}
\hspace{.0\textwidth} \caption{The phase diagrams for the
two-plus-one-flavor NJL model with varying $G_V/G_S$ and fixed
$G_D/G_S=0.95$. Electric-charge-neutrality is considered,  and
only phase diagrams with multiple chiral critical points are
shown. } \label{fig:pd-3flavor}
\end{figure}

\subsection{Model}

In this part, for simplicity, a local two-plus-one-flavor NJL
model is adopted. The two-plus-one-flavor NJL model was developed
in the mid-1980s \cite{3-NJL1,3-NJL2,3-NJL3},  and the most
popular version includes a chiral symmetric four-quark interaction
term and a determination term \cite{KMT} in flavor space
\cite{3-NJL4,3-NJL5,Rehberg:1995kh}. To compare with the previous
study, we take the same model parameters as in
Ref.\cite{Ruester:2005jc} by including the vector interaction
channel. The corresponding Lagrangian density is given by

\begin{eqnarray}
\mathcal{L} &=& \bar \psi \, ( i \dirac - \hat{m} \, ) \psi +G_S
\sum_{i=0}^8 \left[ \left( \bar \psi \lambda_i \psi \right)^2 +
\left( \bar \psi i \gamma_5 \lambda_i \psi \right)^2
\right]-G_V\sum_{i=0}^8 \left[ \left( \bar \psi \gamma^\mu
\lambda_i \psi \right)^2 + \left( \bar \psi i \gamma^\mu \gamma_5
\lambda_i \psi \right)^2 \right]
\nonumber \\
&+& G_D \sum_{\gamma,c} \left[\bar{\psi}_{\alpha}^{a} i \gamma_5
\epsilon^{\alpha \beta \gamma} \epsilon_{abc} (\psi_C)_{\beta}^{b}
\right] \left[ (\bar{\psi}_C)_{\rho}^{r} i \gamma_5 \epsilon^{\rho
\sigma \gamma} \epsilon_{rsc} \psi_{\sigma}^{s} \right]-K \left\{
\det_{f}\left[ \bar \psi \left( 1 + \gamma_5 \right) \psi \right]
+ \det_{f}\left[ \bar \psi \left( 1 - \gamma_5 \right) \psi
\right] \right\} \;, \label{Lagrangian2}
\end{eqnarray}
where the quark spinor field $\psi_{\alpha}^{a}$ carries color
($a=r,g,b$) and flavor ($\alpha=u,d,s$) indices. In contrast to
the two-flavor case, the matrix of the quark current masses is
given by $\hat{m} =\text{diag}_{f}(m_u,m_d,m_s)$ and the Pauli
matrices in flavor space are replaced by the Gell-Mann matrices
$\lambda_i$ in flavor space with $i=1,\ldots,8$, and
$\lambda_0\equiv \sqrt{2/3} \,\openone_{f}$. The corresponding
parametrization of the model parameters is given in
Table(\ref{tab2}), where $G_S$, the coupling constant for the
scalar meson channel, and $K$, the coupling constant responsible
for the $U_A(1)$ breaking, or the KMT term\cite{KMT},
 are fixed by the vacuum
physical observables. The other two coupling constants, $G_V$ and
$G_D$,  are still taken as free parameters in the following.

\begin{table}{
\begin{tabular}{c|c|c|c|c|c}
\hline\hline {\quad $m_{u,d}$(MeV) \quad  } &  \quad {\quad
$m_{s}$(MeV) \quad}&  \quad {\quad $G_S\Lambda^2$ \quad}& \quad
\small {\quad $K\Lambda^5$ \quad}&\quad {\quad $\Lambda$ (MeV)}
\quad &{$M_{u,d}$ (MeV)}\\
\hline 5.5 & 140.7 & 1.835 & 12.36 & 602.3 & 367.7 \\
\hline\hline {\quad $f_{\pi}$(MeV) \quad  } & \quad {\quad
$m_{\pi}$(MeV) \quad} & \quad {\quad $m_K$ (MeV) \quad}& \quad
{\quad $m_{\eta^{,}}$(MeV) \quad} & \quad {\quad $m_{\eta}$(MeV)}
\quad& {$M_{s}$ (MeV)}\\
\hline 92.4 & 135 & 497.7 & 957.8 & 514.8  & 549.5\\
\hline\hline
\end{tabular}
\caption{Model parametrization of the two-plus-one-flavor NJL
model.} \label{tab2}}
\end{table}

\subsection{Thermodynamic potential for neutral color superconductivity}

In general, there exist nine possible two-quark condensates for
the two-plus-one-flavor case with the Lagrangian
~(\ref{Lagrangian2}): three chiral condensates $\sigma_\alpha$,
three diquark condensates $\Delta_c$,  and three vector quark
condensates $\rho_\alpha$, where $\alpha$ and $c$ range from 1 to
3, which stand for three flavors and three colors, respectively.
At the mean-field level, the thermodynamic potential for the
two-plus-one-flavor NJL model  including the charge-neutrality
constraints, is
\begin{eqnarray}
\Omega &=& \Omega_{l} + \frac{1}{4 G_D} \sum_{c=1}^{3} \left|
\Delta_c \right|^2-2 G_V \sum_{\alpha=1}^{3} \rho_\alpha^2
+2 G_S \sum_{\alpha=1}^{3} \sigma_\alpha^2 \nonumber\\
&-& 4 K \sigma_u \sigma_d \sigma_s -\frac{T}{2V} \sum_K \ln \det
\frac{S^{-1}_{MF}}{T} \; , \label{Omega2}
\end{eqnarray}
where $\Omega_{l}$ stands for the contribution from free leptons.
Note that,  for consistency, $\Omega_{l}$ should include the
contributions from both electrons and muons. Since $M_\mu>>M_e$
and $M_e\approx 0$, ignoring the contribution from  muons has
little effect on the phase structure. Therefore, we use
$\Omega_{l}$ given by the last term in Eq.~(\ref{omega-L}).

It should be stressed here that,  for simplicity, the
contributions of the cubic mixing terms among  three different
condensates, such as $\sigma\Delta^2$, $\rho\Delta^2$, and
$\sigma\rho^2$,  are neglected in Eq.~(\ref{Omega2}). These terms
arise from the KMT interaction which may or may not affect the
phase structure. In particular  it was argued in
~\cite{Hatsuda:2006ps} that the cubic mixing term between chiral
 and diquark condensates  may play an important role in the
chiral phase transition in the low temperature region. Beside the
direct contribution of these cubic terms to the thermodynamic
potential, the flavor mixing terms arising from the KMT
interaction also have influence on the dispersion relationship of
quasiquarks. For example, both the diquark condensate and the
quark number density  contribute to the dynamical quark mass.
Therefore, it is a very interesting topic to investigate the
possible effect of these cubic coupling terms on the phase diagram
by using dynamic models of QCD.  Leaving the discussion of this
interesting problem to our future work, here we just simply assume
that none of these mixing terms makes a qualitative difference in
the phase diagram.

Because of the large mass disparity between the strange quark and
the u (d) quark , the favored phase at low temperature and
moderate density might be 2CSC phase rather than CFL as
demonstrated in the two-plus-one-flavor NJL
model\cite{Ruester:2005jc, Abuki:2005ms}. These studies suggest
that the strange quark mass is close to or even larger than $\mu$
near the chiral boundary, which means that the strange quark
density is considerably smaller than that of the u and d quarks.
Therefore, for the two-plus-one-flavor case, the light quarks
still play the dominant role around the chiral boundary,  and the
density mismatch between u and d quarks under the
electric-charge-neutrality constraint is still similar to the
two-flavor case. Since the main purpose of our study is to
investigate the influence of the neutral CSC phase on the chiral
phase transition by taking into account the vector interaction, we
only consider the 2CSC phase in the following.

The inverse quark propagator in the 2CSC phase in the
two-plus-one-flavor case still takes the same form as
$S_{MF}^{-1}$,  with the extended matrixes $\hat{\mu}$ and
$\hat{M}$ in three-flavor space. The constituent quark mass is
given by
\begin{equation}
M_\alpha = m_\alpha - 4 G_S \sigma_\alpha + 2 K \sigma_\beta
\sigma_\gamma \; , \label{Mi}
\end{equation}
and the effective quark chemical potentials take the form
\begin{eqnarray}
\tilde{\mu}_u = \mu - 4 G_V \rho_u-\frac{2}{3}\mu_e, \label{ui}\\
\tilde{\mu}_d = \mu - 4 G_V \rho_d+\frac{1}{3}\mu_e, \label{di}\\
\tilde{\mu}_s = \mu - 4 G_V \rho_s+\frac{1}{3}\mu_e. \label{si}
\end{eqnarray}
The quantity $\bar{\tilde{\mu}}$ ( $\delta\tilde{\mu}$ ) still has
the same form as Eq.(\ref{Average}) [Eq.(\ref{Mismatch})] with
$g(p)\equiv 1$. Ignoring the mass difference between the u quark
and the d quark (the mass difference is very small
\cite{Ruester:2005jc}), the last term in Eq. (\ref{Omega2}) has an
analytical form which greatly simplifies the numerical
calculation. Adopting the variational method, we get the eight
nonlinear coupling equations
\begin{equation}
 \frac{\partial\Omega}{\partial\sigma_u}=
 \frac{\partial\Omega}{\partial\sigma_s}=
 \frac{\partial\Omega}{\partial\Delta}=
 \frac{\partial\Omega}{\partial\rho_{u}}=
 \frac{\partial\Omega}{\partial\rho_{d}}=
\frac{\partial\Omega}{\partial\rho_{s}}=
 \frac{\partial\Omega}{\partial\mu_e}=
 \frac{\partial\Omega}{\partial{\mu_8}}=0\, .
 \label{gapeq}
\end{equation}
Since $\mu_8$ is near zero around the chiral transition region
\cite{Ruester:2005jc,Abuki:2005ms}, taking it to zero makes little
difference in the calculational result and the nonlinear equations
can be reduced to 7.

\subsection{Numerical calculation and discussion}

Similar to the two-flavor case, the vector interaction $G_V/G_S$
is also taken as a free parameter in the following, and the
diquark coupling $G_D/G_S$ is fixed to the standard value. Because
of the contribution from the KMT interaction, the ratio $G_D/G_S$
from Fierz transformation should be 0.95 rather than 0.75 when
only considering the  four-quark interaction \cite{BuballaReview}.
Note that in this case the effective four-quark interaction which
determines the quark constituent mass in vacuum is
$G_{S'}=G_S-\frac{1}{2}K\sigma_s$,  and the standard value of
$G_D/G_S'$ should be 0.75.

The phase diagrams with a multiple critical-point structure for
different vector interactions are shown in
Fig.\ref{fig:pd-3flavor}. One can see that these phase diagrams
are very similar to Fig.\ref{fig:pdset2}, the two-flavor case with
parameter set 2. This is reasonable since these two models almost
reproduce the same constituent quark (u and d) masses and have
similar scale parameters. Figure\ref{fig:pd-3flavor} indicates
that the KMT interaction does not change the possible multiple
critical-point structures for the chiral phase transition.

For simplicity, the unstable regions with chromomagnetic
instability are not plotted in Fig.\ref{fig:pd-3flavor}. The
calculation of the Meissner mass squared in  the 2CSC phase for
the two-plus-one-flavor case is straightforward but complicated.
Including the s quark should have little effect on the value of
the Meissner masses calculated according to the formula for the
two-flavor case \cite{Huang:2004bg} since the s quark does not
take part in Cooper pairing. We can expect that the critical
points E, F,  and G should still be free from the chromomagnetic
instability,  as in Figs.\ref{fig:pdset1} and \ref{fig:pdset2},
since the large strange quark may give a positive contribution
rather than a negative one to the Meissner masses squared. As for
critical point H, it may be located in the unstable region and
could be safe from the instability because the relatively large
$G_V/G_S$ may suppress the unstable region to lower $T$ and higher
$\mu$.

Usually,  it is argued that the instantons should be screened at
large chemical potential and temperature. Therefore, compared to
it's vacuum value, the coupling constant $K$ is expected to be
reduced around the chiral boundary. For smaller $K$, the flavor
mixing effect is suppressed and the mass mismatch between the s
quark and the u(d) quark becomes larger. Accordingly, the
influence of the s quark on the chiral restoration is weakened,
and the situation approaches the two-flavor case. On the other
hand, with decreasing $K$, the u(d) quark mass also decreases
since the contribution from the s quark mass is reduced. This
means the first-order chiral restoration will be weakened when
decreasing $K$. Correspondingly, the COE region should be more
easily formed with the influence of the vector interaction and the
neutral CSC phase , which favors the multiple critical-point
structures or crossover for chiral restoration at low temperature.

Of course, the produced u(d) quark vacuum constituent masses with
different model parameters of the two-plus-one-flavor NJL model
may range from 300-400 MeV, which are all phenomenologically
acceptable just as the two-flavor case. Then, one can expect that
all the critical-point structures found in Sec. \ref{sec: Two
flavor } should also appear in the two-plus-one-flavor case,  even
considering the axial anomaly interaction term.


\section{CONCLUSIONS AND OUTLOOK}

  In this paper, we have explored the effect of
the repulsive vector-vector interaction combined with
electric-charge neutrality in $\beta$ equilibrium  on the chiral
phase and CSC phase transitions within both two-flavor and
two-plus-one-flavor NJL models.

  For the two-flavor case with the presence of the neutral CSC phase, we
demonstrated that,  with the help of the repulsive vector
interaction, there always exists a parameter window in the NJL
models which favors the appearance of a multiple chiral
critical-point structure for a wide range of the vacuum quark
mass, i.e.,
 from 300 MeV to 400 MeV.
Besides the two- and three-critical-points structures found in
\cite{KitazawaVector,Zhang:2008wx} and  \cite{Zhang:2008wx},
respectively, we have shown for the first time that
 a four-critical-points
structure of the QCD phase diagram can be realized
; such a multiple critical-point structure is caused by  the joint
effect of positive $\mu_e$ and $G_V$. Because the dynamical
strange quark mass is still relatively large near the boundary of
the chiral transition, the multiple critical-point structures
present in the two-flavor case also appear in the
two-plus-one-flavor case. For the intermediate diquark coupling
case,  the number of critical points changes  as $1\, \rightarrow
\, 2\,\rightarrow\, 4\,\rightarrow\, 2\, \rightarrow\,0$ with an
increasing vector coupling
 in the two-plus-one-flavor NJL model.
In general, one can expect that different model parameters
may possibly give other
 order of the number of the critical points
as the  vector coupling is increased.

Although our analysis
is based on a low-energy effective  model which inherently has,
more or less,  a parameter dependence, we have seen that the
physical mechanism to realize the multiple critical-point
structure is solely dependent on the basic ingredients of the
effective quark dynamics and thermodynamics. Therefore, we believe
that the results obtained in the present work
 should be taken seriously and examined in other effective models of
 QCD,  or hopefully lattice QCD simulations.
Our result also has a meaningful implication for the study of
phase transitions in condensed matter physics. That means some
external constraints enforced on the system can lead to the
formation or expansion of the coexisting phase,  and the
competition between two order parameters can give rise to multiple
critical points.

Last but not least, we emphasize that we have shown for the first
time  that the  repulsive vector interaction which should
generically exist between the quarks does suppress the
chromomagnetic instability related to the asymmetric homogeneous
2CSC phase.
 With increasing vector interaction,
the unstable region associated with chromomagnetic instability
shrinks towards lower temperatures and higher chemical potentials.
This means that the vector interaction can partially or even
totally resolve  the chromomagnetic instability problem.

Note that to cure the chromomagnetic instability, inhomogeneous
asymmetric color superconductivity phases such as the LOFF phase
and the gluonic phase were proposed in the literature. For the
inhomogeneous phase, beside the condensate
$\langle\overline{\psi}{\gamma_0}\psi\rangle$, there is no reason
to rule out the appearance of another new condensate ,
$\langle\overline{\psi}\vec{\gamma}\psi\rangle$,  when considering
the vector interaction. The effect of both the timelike vector
condensate and the spacelike condensate on the asymmetric
inhomogeneous  CSC phase will be reported in our future
work\cite{future}.

\acknowledgments

 We acknowledge Kenji Fukushima for useful discussion.
One of the authors ( Z.~Z. ) is grateful for the support from the
Grants-in-Aid provided by Japan Society for the Promotion of
Science (JSPS). This work was partially supported by a
Grant-in-Aid for Scientific Research by the Ministry of Education,
Culture, Sports, Science and Technology (MEXT) of Japan (No.
20540265 and No. 19$\cdot$07797),
 by Yukawa International Program for Quark-Hadron Sciences, and by the
Grant-in-Aid for the global COE program `` The Next Generation of
Physics, Spun from Universality and Emergence '' from MEXT.


\begin{thebibliography}{99}


\bibitem{Asakawa:1989bq}
  M.~Asakawa and K.~Yazaki,
  Nucl.\ Phys.\  A {\bf 504}, 668 (1989).

\bibitem{Barducci:1989}
 A.~Barducci, R.~Casalbuoni, S.~De Curtis, R.~Gatto and G.~Pettini,
  Phys.\ Lett.\  B {\bf 231}, 463 (1989);  \,
 Phys.\ Rev.\  D {\bf 49}, 426 (1994).

\bibitem{Stephanov:2007fk}
As a review, see, M.~A.~Stephanov,
  Prog.\ Theor.\ Phys.\ Suppl.\  {\bf 153}, 139 (2004)
  [Int.\ J.\ Mod.\ Phys.\  A {\bf 20}, 4387 (2005)];\,
PoS {\bf LAT2006}, 024 (2006)  [arXiv:hep-lat/0701002].

\bibitem{WilczekReview}
  K.~Rajagopal and F.~Wilczek,
  arXiv:hep-ph/0011333.

\bibitem{RischkeReview}
  D.~H.~Rischke,
  Prog.\ Part.\ Nucl.\ Phys.\  {\bf 52}, 197 (2004)
  [arXiv:nucl-th/0305030].

\bibitem{BuballaReview}
  M.~Buballa,
  Phys.\ Rept.\  {\bf 407}, 205 (2005)
  [arXiv:hep-ph/0402234].

\bibitem{AlfordReview}
  M.~G.~Alford, A.~Schmitt, K.~Rajagopal and T.~Schafer,
  arXiv:0709.4635 [hep-ph].

\bibitem{CFL}
  M.~G.~Alford, K.~Rajagopal and F.~Wilczek,
  Nucl.\ Phys.\  B {\bf 537}, 443 (1999)
  [arXiv:hep-ph/9804403].
%
\bibitem{KitazawaVector}
  M.~Kitazawa, T.~Koide, T.~Kunihiro and Y.~Nemoto,
  Prog.\ Theor.\ Phys.\  {\bf 108}, 929 (2002)
  [arXiv:hep-ph/0207255].
%

\bibitem{Klimt}
  S.~Klimt, M.~Lutz and W.~Weise,
  Phys.\ Lett.\  B {\bf 249}, 386 (1990).

\bibitem{Buballa1996}
  M.~Buballa,
  Nucl.\ Phys.\  A {\bf 611}, 393 (1996)
  [arXiv:nucl-th/9609044].


\bibitem{ref:EHS}
N.~Evans, S.~D.~H.~Hsu and M.~Schwetz, Nucl.~Phys.~ B {\bf 551},
275 (1999).

\bibitem{ref:SW-reno}
T.~Sch\"afer and F.~Wilczek, Phys.~Lett.  B {\bf 450}, 325 (1999).

\bibitem{ref:SS}
T.~Sch\"afer and E.~Shuryak, Rev.~Mod.~Phys. {\bf 70}, 323 (1998).

\bibitem{ref:RWP}
C. D. Roberts and A. G. Williams, Prog. Part. Nucl. Phys. 33
(1994) 477; P. C. Tandy, Prog. Part. Nucl. Phys. 39 (1997) 117.

\bibitem{Ebert:1983}
 D.~Ebert and M.~K.~Volkov,
  Z.\ Phys.\  C {\bf 16}, 205 (1983).

\bibitem{3-NJL1}
D.~Ebert and H.~Reinhardt,
  Nucl.\ Phys.\  B {\bf 271}, 188 (1986).

\bibitem{Klimt:1989pm}
 M.~Takizawa, K.~Tsushima, Y.~Kohyama and K.~Kubodera,
  Prog.\ Theor.\ Phys.\  {\bf 82}, 481 (1989); \\
  S.~Klimt, M.~Lutz, U.~Vogl and W.~Weise,
  Nucl.\ Phys.\  A {\bf 516}, 429 (1990).

\bibitem{Hatsuda:2006ps}
  T.~Hatsuda, M.~Tachibana, N.~Yamamoto and G.~Baym,
  Phys.\ Rev.\ Lett.\  {\bf 97}, 122001 (2006)
  [arXiv:hep-ph/0605018];
  Phys.\ Rev.\  D {\bf 76}, 074001 (2007)
  [arXiv:0704.2654 [hep-ph]].

\bibitem{Continuity}
  T.~Schafer and F.~Wilczek,
  Phys.\ Rev.\ Lett.\  {\bf 82}, 3956 (1999)
  [arXiv:hep-ph/9811473];
  M.~G.~Alford, J.~Berges and K.~Rajagopal,
  Nucl.\ Phys.\  B {\bf 558}, 219 (1999)
  [arXiv:hep-ph/9903502].


\bibitem{Zhang:2008wx}
  Z.~Zhang, K.~Fukushima and T.~Kunihiro,
  Phys.\ Rev.\  D {\bf 79}, 014004 (2009)
  [arXiv:0808.3371 [hep-ph]].

\bibitem{Huang:2004bg}
  M.~Huang and I.~A.~Shovkovy,
  Phys.\ Rev.\  D {\bf 70}, 051501(R) (2004)
  [arXiv:hep-ph/0407049];
  \textit{ibid.} D {\bf 70}, 094030 (2004)
  [arXiv:hep-ph/0408268].


\bibitem{Ruester:2005jc}
  S.~B.~Ruester, V.~Werth, M.~Buballa, I.~A.~Shovkovy and D.~H.~Rischke,
  Phys.\ Rev.\  D {\bf 72}, 034004 (2005)
  [arXiv:hep-ph/0503184].

\bibitem{Abuki:2005ms}
  H.~Abuki and T.~Kunihiro,
  Nucl.\ Phys.\  A {\bf 768}, 118 (2006)
  [arXiv:hep-ph/0509172].

\bibitem{KMT} M.~Kobayashi and T.~Maskawa, Prog.~ Theor.~Phys. {\bf 44}, 1422 (1970);\,
M.~Kobayashi, H.~Kondo and T.~Maskawa, Prog.~ Theor.~Phys. {\bf 45}, 1955 (1971);\\
G.~'t Hooft, Phys. Rev. {\bf D14}, 3432 (1976); Phys. Rep. {\bf 142}, 357 (1986).


\bibitem{Hatsuda:1994}
  T.~Hatsuda and T.~Kunihiro,
 Phys.\ Rev.\ Lett.\  {\bf 55}, 158 (1985);\,
  Phys.\ Rept.\  {\bf 247}, 221 (1994)
  [arXiv:hep-ph/9401310].

\bibitem{Klevansky:1992}
  S.~P.~Klevansky,
  Rev.\ Mod.\ Phys.\  {\bf 64}, 649 (1992).


\bibitem{Alford:1997zt}
  M.~G.~Alford, K.~Rajagopal and F.~Wilczek,
  Phys.\ Lett.\  B {\bf 422}, 247 (1998)
  [arXiv:hep-ph/9711395].

\bibitem{Schmidt:1994di}
  S.~M.~Schmidt, D.~Blaschke and Yu.~L.~Kalinovsky,
  Phys.\ Rev.\  C {\bf 50}, 435 (1994).
\bibitem{Bowler:1994ir}
  R.~D.~Bowler and M.~C.~Birse,
  Nucl.\ Phys.\  A {\bf 582}, 655 (1995)
  [arXiv:hep-ph/9407336].
\bibitem{Blaschke:2000gd}
  D.~Blaschke, G.~Burau, Yu.~L.~Kalinovsky, P.~Maris and P.~C.~Tandy,
  Int.\ J.\ Mod.\ Phys.\  A {\bf 16}, 2267 (2001)
  [arXiv:nucl-th/0002024].
\bibitem{GomezDumm:2005hy}
  D.~Gomez Dumm, D.~B.~Blaschke, A.~G.~Grunfeld and N.~N.~Scoccola,
  Phys.\ Rev.\  D {\bf 73}, 114019 (2006)
  [arXiv:hep-ph/0512218].
\bibitem{Aguilera:2006cj}
  D.~N.~Aguilera, D.~Blaschke, H.~Grigorian and N.~N.~Scoccola,
  Phys.\ Rev.\  D {\bf 74}, 114005 (2006)
  [arXiv:hep-ph/0604196].

\bibitem{Grigorian:2006qe}
  H.~Grigorian,
  Phys.\ Part.\ Nucl.\ Lett.\  {\bf 4}, 223 (2007)
  [arXiv:hep-ph/0602238].

\bibitem{Kiriyama:2006jp}
  O.~Kiriyama,
  Phys.\ Rev.\  D {\bf 74}, 114011 (2006)
  [arXiv:hep-ph/0609185]


\bibitem{Alford:2002kj}
  M.~Alford and K.~Rajagopal,
  JHEP {\bf 0206}, 031 (2002)
  [arXiv:hep-ph/0204001];
  A.~W.~Steiner, S.~Reddy and M.~Prakash,
  Phys.\ Rev.\  D {\bf 66}, 094007 (2002)
  [arXiv:hep-ph/0205201].

\bibitem{Giannakis:2004pf}
  I.~Giannakis and H.~C.~Ren,
  Phys.\ Lett.\  B {\bf 611}, 137 (2005)
  [arXiv:hep-ph/0412015].

\bibitem{Gorbar:2005rx}
  E.~V.~Gorbar, M.~Hashimoto and V.~A.~Miransky,
  Phys.\ Lett.\  B {\bf 632}, 305 (2006)
  [arXiv:hep-ph/0507303];
  Phys.\ Rev.\ Lett.\  {\bf 96}, 022005 (2006)
  [arXiv:hep-ph/0509334];
  Phys.\ Rev.\  D {\bf 75}, 085012 (2007)
  [arXiv:hep-ph/0701211].


\bibitem{Kiriyama:2006ui}
  O.~Kiriyama, D.~H.~Rischke and I.~A.~Shovkovy,
  Phys.\ Lett.\  B {\bf 643}, 331 (2006)
  [arXiv:hep-ph/0606030];

\bibitem{He:2007cn}
L.~He, M.~Jin and P.~Zhuang,
  Phys.\ Rev.\  D {\bf 75}, 036003 (2007)
  [arXiv:hep-ph/0610121].

\bibitem{Fukushima:2005cm}
  K.~Fukushima,
  Phys.\ Rev.\  D {\bf 72}, 074002 (2005)
  [arXiv:hep-ph/0506080].


\bibitem{Kitazawa:2006zp}
  M.~Kitazawa, D.~H.~Rischke and I.~A.~Shovkovy,
  Phys.\ Lett.\  B {\bf 637}, 367 (2006)
  [arXiv:hep-ph/0602065].


\bibitem{3-NJL2}
 V.~Bernard, R. L. Jaffe and U. G. ~Meissner, Phys.\ Lett.\  B {\bf 198}, 92 (1987).

\bibitem{3-NJL3}
 T.~Hatsuda and T.~Kunihiro, Phys.\ Lett.\  B {\bf 198}, 126 (1987).

\bibitem{3-NJL4}
T.~Kunihiro and T.~Hatsuda, Phys.\ Lett.\  B {\bf 206}, 385 (1988); Erratum-ibid. {\bf 210}, 278 (1988);\\
T.~Kunihiro, Phys.\ Lett.\  B {\bf 219}, 363 (1989).

\bibitem{3-NJL5}
 V.~Bernard, R. L. Jaffe and U. G. ~Meissner, Nucl.\ Phys.\  B {\bf 308}, 753
 (1988).

\bibitem{Rehberg:1995kh}
  P.~Rehberg, S.~P.~Klevansky and J.~Hufner,
  Phys.\ Rev.\  C {\bf 53}, 410 (1996)
  [arXiv:hep-ph/9506436].

\bibitem{future}
 Z.~Zhang and T.~Kunihiro( unpublished ).


\end{thebibliography}
\end{document}